\documentclass[12pt,a4paper]{article}
\usepackage[includeheadfoot,
marginratio={1:1,2:3},
width=551pt,
height=786pt,]{geometry}

\usepackage{amsfonts}
\usepackage{amssymb}
\usepackage{comment}
\usepackage{graphicx}
\usepackage{amsmath}
\usepackage{tabu}
\usepackage{dsfont}
\usepackage{float}
\usepackage{amsthm}
\usepackage{slashed}
\usepackage{setspace}
\usepackage[labelformat=simple]{subcaption}

\usepackage{pgfplots}
\usepackage{tikz}
\usepackage{tikz-cd}
\usetikzlibrary{shapes.misc,shadows,fit,decorations.pathmorphing,positioning,trees,decorations.markings,decorations.pathreplacing,calc,shapes,patterns,arrows,positioning,arrows.meta}
\usepackage{xcolor}
\usepackage{pgfplots}
\usepackage{pgfplotstable}
\usepackage{xcolor}
\usepackage[boxsize=0.5em,aligntableaux=center]{ytableau}
\usepackage{makecell}
\usepackage{diagbox}
\usepackage{environ}

\usepackage{amsmath, amsthm, amssymb, amsfonts}
\usepackage{amsmath, amsthm, amssymb, amsfonts}
\usepackage{multirow}
\usepackage{graphicx}
\usepackage{float}
\usepackage[numbers,sort&compress]{natbib}
\usepackage{enumerate}
\usepackage{enumitem}
\usepackage{hhline}
\usepackage{mathtools}
\usepackage{longtable}
\usepackage{alphalph}

\usepackage{amsfonts}
\usepackage{amssymb}
\usepackage{comment}
\usepackage{graphicx}
\usepackage{amsmath}
\usepackage{tabu}
\usepackage{dsfont}
\usepackage{float}
\usepackage{amsthm}
\usepackage{slashed}
\usepackage{setspace}
\usepackage[labelformat=simple]{subcaption}

\usepackage[boxsize=0.5em,aligntableaux=center]{ytableau}

\usepackage{amsmath, amsthm, amssymb, amsfonts}
\usepackage[font=small, labelfont=bf]{caption}
\usepackage{amsmath, amsthm, amssymb, amsfonts}
\usepackage{multirow}
\usepackage{graphicx}
\usepackage{float}
\usepackage[numbers,sort&compress]{natbib}

\usepackage{enumerate}

\usepackage{amsmath}
\usepackage{amsfonts}
\usepackage{amssymb}
\usepackage{graphicx}
\usepackage{cancel}
\usepackage{empheq}
\usepackage{color}
\usepackage{hyperref}

\usepackage{float}
\usepackage{framed}
\restylefloat{table}
\restylefloat{figure}

\newcommand{\nc}{\newcommand}
\nc{\beq}{\begin{equation}}
\nc{\eeq}{\end{equation}}
\nc{\bea}{\begin{eqnarray}}
\nc{\eea}{\end{eqnarray}}

\allowdisplaybreaks[2]
\numberwithin{equation}{section}

\def\rmt{\rm t}
\def\rmz{\rm z}

\begin{document}

	{\hfill
		%
	
	\vspace{1.0cm}
	\begin{center}
		{\Large
			On stable type IIA de-Sitter vacua with geometric flux}
		\vspace{0.4cm}
	\end{center}
	
	\vspace{0.35cm}
	\begin{center}
		Pramod Shukla\footnote{Email: pshukla@jcbose.ac.in}
	\end{center}

	\vspace{0.1cm}
	\begin{center}
		{ICTP, Strada Costiera 11, Trieste 34151, Italy; \\ \vskip0.15in
		Department of Physics, University of Allahabad, Prayagraj 211002, India; \\ \vskip0.15in
		Department of Physics, Unified Academic Campus, Bose Institute, \\ EN 80, Sector V, Bidhannagar, Kolkata 700 091, India.}
	\end{center}
	
	\vspace{1cm}
	
\abstract{We present a class of tachyon-free (stable) AdS as well as dS solutions in the context of type IIA orientifold compactifications. This model is based on a type IIA setting having geometric flux  which, in addition, also includes the usual NS-NS three-form flux ${\rm H}_3$ and the RR $p$-form fluxes ${\rm F}_p$ for $p \in \{0, 2, 4, 6\}$. Although all the saxionic moduli are stabilized, there remains a combination of the RR axions which is flat, and the dS solution is realized through the $D$-term effects induced by the geometric fluxes, without which one can only have AdS solutions. Using flux scaling arguments we also discuss how to engineer a parametric control in these models, in the sense of realizing the respective AdS/dS solutions in large volume, weak string-coupling as well as large complex-structure regime. We also discuss open possibilities (e.g. about the unknown status on having the complete set of Bianchi identities) under which such tachyon-free dS solutions may or may not be viable.}

\clearpage
	
\tableofcontents


\section{Introduction}
\label{sec_intro}
Realizing de-Sitter solutions and possible obstructions on the way of doing it have been always in the center of attraction since decades. In this regard, the historicity of efforts on finding de-Sitter vacua in string theory (inspired) setups can be classified into the following three categories: 
\begin{itemize}
\item{Existence: Starting from a simple model building with minimal ingredients at hand, there have been several de-Sitter no-go scenarios proposed from time to time. For example, several no-go theorems forbidding de-Sitter (dS) and inflationary realizations have been proposed in a series of works \cite{Maldacena:2000mw, Hertzberg:2007wc, Hertzberg:2007ke, Haque:2008jz, Flauger:2008ad, Caviezel:2008tf,  Covi:2008ea, deCarlos:2009fq, Caviezel:2009tu, Danielsson:2009ff, Danielsson:2010bc, Wrase:2010ew,  Shiu:2011zt, McOrist:2012yc, Dasgupta:2014pma, Gautason:2015tig, Junghans:2016uvg, Andriot:2016xvq, Andriot:2017jhf, Danielsson:2018ztv,Shukla:2019dqd,Shukla:2019akv,Marchesano:2020uqz}, mostly in the context of type IIA orientifold compactifications. In fact these no-go results have played a central role in the recent revival of the swampland conjectures \cite{Ooguri:2006in, Obied:2018sgi}. However, the good thing about any no-go result is the fact that it happens to hold in a framework with a given set of assumptions, and for a model builder aiming to realize de-Sitter solution, the very first task is to explore the loopholes or the limits under which the no-go necessities can be evaded, e.g. see \cite{Andriot:2022way} for a recent charting of the (anti-) de Sitter vacua from the perspective of ten-dimensional supergravities. In contrary to the (minimal) de-Sitter no-go scenarios, in the meantime there have been several proposals for realizing (stable) de-Sitter vacua \cite{Kachru:2003aw, Burgess:2003ic, Achucarro:2006zf, Westphal:2006tn, Silverstein:2007ac, Rummel:2011cd, Cicoli:2012fh, Louis:2012nb, Cicoli:2013cha, Cicoli:2015ylx, Cicoli:2017shd, Akrami:2018ylq, Antoniadis:2018hqy,Antoniadis:2018ngr, Antoniadis:2019rkh, Basiouris:2020jgp, Antoniadis:2020stf, Cicoli:2018kdo, Crino:2020qwk, Cicoli:2021dhg}; see \cite{Heckman:2019dsj, Heckman:2018mxl} also for the $F$-theoretic initiatives taken in this regard.}

\item{Stability: After evading some no-go results by adding more ingredients, the immediate question is about the stability of the subsequent de-Sitter solution, especially to ensure if those solutions are true minima or tachyonic in nature. This is a quite crucial condition to check as there have been plethora of de-Sitter solutions constructed by simple attempts of evading the no-go results, many of which eventually turned out to be tachyonic in nature \cite{Flauger:2008ad,Hertzberg:2007ke, Haque:2008jz, Danielsson:2009ff, Danielsson:2011au, Chen:2011ac, Danielsson:2012et,Junghans:2016uvg}. In fact the existence of such tachyonic de-Sitter solutions demanded the need of refinements in the original swampland conjectures \cite{Ooguri:2006in, Obied:2018sgi} leading to numerous amount of work with similar interests including the Quintessence alternative \cite{Garg:2018reu, Agrawal:2018own, Andriot:2018wzk, Andriot:2018ept, Denef:2018etk, Conlon:2018eyr, Roupec:2018mbn, Murayama:2018lie, Choi:2018rze, Hamaguchi:2018vtv, Olguin-Tejo:2018pfq, Blanco-Pillado:2018xyn,Lin:2018kjm, Han:2018yrk, Raveri:2018ddi, Dasgupta:2018rtp, Danielsson:2018qpa, Andriolo:2018yrz, Dasgupta:2019gcd, Andriot:2019wrs,Palti:2019pca,Cicoli:2012tz,Junghans:2022exo} and the challenges it faces regarding the discrepancy in Hubble constant \cite{Banerjee:2020xcn, Lee:2022cyh}.}

\item{Viability: This step is usually combined with what we referred to `Stability' in the second step. However, given the fact that this appears to cover a wider range of open possibilities, it may be worth considering it separately. Given that very little is known about the corrections which induce the scalar potential pieces to perform moduli stabilization, it is an important question to ask if the de-Sitter vacua which pass the tests in the first two steps are genuine or not. For example, the questions regarding scale separation and field excursions in moduli space \cite{Blumenhagen:2017cxt, Blumenhagen:2018nts, Blumenhagen:2018hsh, Palti:2017elp, Conlon:2016aea, Hebecker:2017lxm, Klaewer:2016kiy, Baume:2016psm, Landete:2018kqf, Cicoli:2018tcq, Font:2019cxq, Grimm:2018cpv, Hebecker:2018fln, Banlaki:2018ayh, Junghans:2018gdb,Junghans:2020acz,Apers:2022zjx} without breaking effective field theory (EFT) assumptions, tadpole conjecture \cite{Bena:2020xrh,Plauschinn:2021hkp,Plauschinn:2020ram} and ways to avoid it \cite{Marchesano:2021gyv} may be considered in this class. In fact, it happens very often that the scalar potential corrections are known only in pieces and are sometimes discovered/challenged with new updates, and in this regard, a perfect  check about viability may be considered as the toughest task for string phenomenologists !}

\end{itemize}

\noindent
In the current work our aim is to analyse the flux vacua arising from a concrete type IIA CY orientifold model along the aforesaid three steps. For that purpose we begin with considering a simple type IIA model which includes geometric flux along with the standard three-form NS-NS $({\rm H}_3)$ flux and the RR $p$-form ${\rm F}_p$ fluxes for $p \in \{0, 2, 4, 6\}$. In our approach we first investigate the de-Sitter scenarios which can be obtained by evading the standard no-go results, e.g. those obtained from the moduli dynamics restricted to the volume/dilaton plane \cite{Haque:2008jz, Caviezel:2008tf, Flauger:2008ad}. In this process we make some specific choice of fluxes such that 
\begin{itemize}

\item
The only geometric flux which arise in the scalar potential corresponds to the $D$-term effects which are positive semi-definite. All the geometric flux contributions contributing to the $F$-term effects in the scalar potential are set to zero, which subsequently also facilitates in satisfying the NS-NS Bianchi identities without loosing any $D$-term fluxes.

\item
We consider only some `suitable' RR flux components to be non-zero while setting many of those to zero. This facilitates in making the tadpole terms independent of one flux which subsequently does not get bounded by $O6$-charge through $D6$ tadpole conditions, and turns out to be useful in realizing larger volume and weaker string coupling after the minimization process. However, recalling the fact that Romans mass term along with geometric flux is needed to evade the previously known no-go results, we always keep non-zero Romans mass along with some geometric flux components to be non-zero.

\end{itemize}

\noindent
The $T$-dual completions of the four-dimensional type II effective theories by including the (non-)geometric fluxes have been initiated in the toroidal context \cite{Aldazabal:2006up, deCarlos:2009fq, deCarlos:2009qm, Aldazabal:2010ef, Lombardo:2016swq, Lombardo:2017yme}, and a couple of interesting efforts have been initiated in establishing a concrete connection between the (non-)geometric ingredients of the two theories in \cite{Grana:2006hr, Benmachiche:2006df}, and a full mapping with the $T$-duality at the level of NS-NS non-geometric flux components and the two scalar potentials, has been presented in \cite{Shukla:2019wfo}. In \cite{Shukla:2019dqd}, it was shown that a geometric type IIA setup corresponds to a non-geometric type IIB under a set of $T$-dual transformations relating the fluxes and moduli of the two theories (e.g. see \cite{Shukla:2019wfo}). In fact many of the (geometric) type IIA de-Sitter no-go scenarios have been $T$-dualized to type IIB case with non-geometric fluxes \cite{Shukla:2019dqd}, which has refined the regime of de-Sitter search in various possible ways. 

However, regarding the viability there remains a subtle and open issue. It is naively assumed that a consistent incorporation of the various (non-)geometric fluxes enriches the compactification backgrounds which creates better possibilities for model building, however one does not known how many and which type of fluxes can be simultaneously turned-on, respecting the various Bianchi identities and the tadpole cancellation conditions. In this regard, there have been two formulations for computing the Bianchi identities; the standard one mostly applicable to toroidal orientifolds involves fluxes with non-cohomology indices \cite{Shelton:2005cf, Ihl:2007ah, Aldazabal:2008zza}, while in the cohomology formulation fluxes are represented using the non-trivial cohomology indices \cite{Ihl:2007ah, Benmachiche:2006df, Grana:2006hr}. However, a mismatch between the two sets of Bianchi identities of these two formulations have been observed in \cite{Ihl:2007ah,Robbins:2007yv} and studied in some good detail in \cite{Shukla:2016xdy,Gao:2018ayp}. The additional unknown identities in the cohomology formulation, if they exist, might be relevant for our de-Sitter solution along with the recent studies performed in \cite{Blumenhagen:2015qda, Blumenhagen:2015kja,Blumenhagen:2015jva, Blumenhagen:2015xpa, Li:2015taa, Betzler:2019kon}.

This article is organised as follows. Section \ref{sec_prelims} presents a brief collection of the relevant ingredients about the generic type IIA models with geometric flux, having some discussion about the possible scenarios in which the standard/known de-Sitter no-go results can be evaded, along with presenting an explicit type IIA construction with geometric flux. Subsequently, a detailed analysis of the AdS as well as dS solutions is presented in section \ref{sec_AdS-dS-vacua}. Finally we summarise our results in section \ref{sec_conclusions} and provide the additional useful pieces of information in appendix \ref{sec_appendix1}.


\section{Type IIA orientifold with geometric flux}
\label{sec_prelims}

The four-dimensional effective potentials arising from type II flux compactifications and their applications towards moduli stabilization have been extensively studied in series of works \cite{Kachru:2003aw, Taylor:1999ii, Blumenhagen:2003vr, Grimm:2004ua, Grimm:2004uq, Denef:2005mm, Grana:2005jc, Balasubramanian:2005zx, Blumenhagen:2006ci, Douglas:2006es, Blumenhagen:2007sm}. In particular, the study of nongeometric flux compactifications and their four-dimensional scalar potentials have led to a continuous progress in various phenomenological aspects such as towards moduli stabilization, in constructing de-Sitter vacua and also in realizing the minimal aspects of inflationary cosmology \cite{Aldazabal:2006up, Ihl:2006pp, Ihl:2007ah, Font:2008vd, Guarino:2008ik, Aldazabal:2008zza, deCarlos:2009qm,  Danielsson:2012by, Blaback:2013ht, Damian:2013dwa, Damian:2013dq, Hassler:2014mla, Blumenhagen:2014gta, Blumenhagen:2015qda,  Blumenhagen:2015jva, Li:2015taa, Blumenhagen:2015kja, Blumenhagen:2015xpa, Blaback:2015zra}. In the context of Type II supergravity theories, such (non-)geometric fluxes can generically induce tree-level contributions to the scalar potential for all the moduli and hence can subsequently help in dynamically stabilizing them through the lowest order effects. Moreover, the common presence of the nongeometric fluxes in Double Field Theory (DFT), superstring flux-compactifications, and the gauged supergravities has helped in understanding a variety of interconnecting aspects in these three formulations along with opening new windows for exploring some phenomenological aspects as well \cite{Derendinger:2004jn, Derendinger:2005ph, Shelton:2005cf, Wecht:2007wu, Aldazabal:2006up, Dall'Agata:2009gv, Aldazabal:2011yz, Aldazabal:2011nj, Geissbuhler:2011mx, Grana:2012rr, Dibitetto:2012rk, Andriot:2013xca, Andriot:2014qla, Blair:2014zba, Andriot:2012an, Geissbuhler:2013uka, Guarino:2008ik, Blumenhagen:2013hva, Villadoro:2005cu, Robbins:2007yv, Lombardo:2016swq, Lombardo:2017yme}. Moreover, the nongeometric flux compactification scenarios also present some interesting utilisations of the symplectic geometries \cite{Ceresole:1995ca, D'Auria:2007ay} to formulate the effective scalar potentials; e.g. see \cite{Shukla:2015hpa, Gao:2017gxk, Shukla:2016hyy}, which generalize the work of \cite{Taylor:1999ii, Blumenhagen:2003vr} by including the nongeometric fluxes. The ten-dimensional origin of the four-dimensional nongeometric scalar potentials have been explored via an iterative series of works in the supergravities \cite{Villadoro:2005cu, Blumenhagen:2013hva, Gao:2015nra, Shukla:2015rua, Shukla:2015bca, Andriot:2012wx, Andriot:2012an,Andriot:2014qla, Blair:2014zba, Andriot:2013xca, Andriot:2011uh, D'Auria:2007ay} and through some robust realization of the Double Field Theory (DFT) reduction on Calabi Yau threefolds \cite{Blumenhagen:2015lta,Plauschinn:2018wbo}. Moreover, a concrete connection among the type II effective potentials derived from DFT reductions and those of the symplectic approach has been established in \cite{Shukla:2015hpa, Gao:2017gxk}. In this section we recollect the relevant ingredients for type IIA Calabi-Yau orientifold models, and more details can be directly refereed to \cite{Shukla:2019wfo}. 

\subsection{A couple of scenarios evading the known dS no-go results}
The moduli dynamics in the low energy four-dimensional effective supergravity theories are governed by the so-called F- and D-term contributions which are encoded in the K\"ahler potential, the (flux) superpotential and the gauge kinetic functions. The four-dimensional geometric type IIA scalar potential arising from these three ingredients can be also collected as under \footnote{For completeness, we present the generic scalar potential (\ref{eq:pot2}) and its simpler formulations in the appendix \ref{sec_appendix1}.},
\bea
\label{eq:gpot1}
& & V  \equiv 
\left(V_{f_6} + V_{f_4} + V_{f_2} + V_{f_0}\right) + \left(V_{h} + V_{\omega}\right) + V_{\rm loc}, 
\eea
where
\bea
\label{eq:typeIIA-genpot3}
& & V_{f_0} = \, {e^{4D}\, \rho^3}\, A_{1}\,, \quad V_{f_2} =  \, {e^{4D}\, \rho}\, A_{2}\,, \quad V_{f_4} = \, \frac{e^{4D}}{\rho} \, A_{3}\,, \quad V_{f_6} = \, \frac{e^{4D}}{\rho^3} \,A_{4}\,, \\
& & V_h = \frac{e^{2D}}{\rho^3 \,\sigma^3} \, \left(A_5 + \sigma^2 \, A_6+ \sigma^4 \, A_7 \right)\,, \quad V_\omega = \frac{e^{2D}}{\rho\,\sigma^3} \, \left(A_8 + \sigma^2 \, A_9+ \sigma^4 \, A_{10} + \sigma^6 \, A_{11} \right) \,,\nonumber\\
& & V_{\rm loc} = \, \frac{e^{3D}}{\sigma^\frac32} \,\left(A_{12} + A_{13} \, \sigma^2\right). \nonumber
\eea  
Here, apart from having the 4-dimensional dilaton modulus $D$, we have introduced two other moduli, namely $\rho$ and $\sigma$, via the redefinition in the overall volume (${\cal V}$) of the Calabi-Yau threefold and its mirror volume ${\cal U}$ by considering their respective two-cycle volume moduli $\rmt^a$ and $\rmz^i$ as below, 
\bea
& & {\rmt}^a = \rho \, \gamma^a, \quad \implies \quad {\cal V} = \rho^3, \qquad \kappa_{abc}\gamma^a\gamma^b\gamma^c = 6,\\
& & {\rmz}^\lambda = \sigma \, \theta^\lambda, \quad \implies \quad {\cal U} = \sigma^3, \qquad k_{\rho\gamma\delta} \theta^\rho\theta^\gamma\theta^\delta = 6,\nonumber
\eea
where $\gamma^a$'s denote some angular K\"ahler moduli corresponding to the compactifying Calabi-Yau threefold while $\theta^\lambda$'s corresponds to the angular K\"ahler moduli on the mirror Calabi-Yau threefold. Similarly, $\kappa_{abc}$ denotes triple-intersection numbers on the Calabi-Yau threefold while $k_{\rho\gamma\delta}$ denotes the same for the respective mirror Calabi-Yau threefolds. In addition, the quantities $A_i$'s in Eqn. (\ref{eq:typeIIA-genpot3}) denote some functions of fluxes, and the moduli other than volume modulus $\rho$, the complex-structure modulus $\sigma$ and the 4-dimensional dilaton $D$. For completeness, the explicit expressions of $A_i$'s are given in Eqn.~(\ref{eq:typeIIA-genpot4}) of the Appendix \ref{sec_appendix1}. In fact one finds that all the $A_i$'s except $A_9,\, A_{10}, \, A_{12}$ and $A_{13}$ are positive semi-definite. 

There have been several no-go results against the de-Sitter realization in type IIA based simple models, especially arising from the volume/dilaton analysis \cite{Hertzberg:2007wc,Flauger:2008ad}. One possible method to evade such no-go results is to include the geometric flux along with a non-zero Romans mass term. Several attempts for de-Sitter realization have been made exploiting this observations \cite{Flauger:2008ad}, {\it however best to our knowledge there is no proposed model which realizes non-tachyonic de-Sitter solution using integer valued fluxes satisfying the Bianchi identities, at least all of the known ones.} Let us note that the scalar potential pieces in Eq. (\ref{eq:typeIIA-genpot3}) have complicated coefficients (depending on other moduli) and therefore some of the well known de-Sitter no-go results arising from minimalistic volume/dilaton analysis, e.g. \cite{Hertzberg:2007wc,Flauger:2008ad,Junghans:2020acz,Apers:2022zjx}, may be easily evaded for the more generic constructions. This scalar potential can be studied for the searching the stable vacua through minimization of the moduli/axions, and in our generic approach we take the following two steps:
\begin{itemize}
\item
Step-1: First we consider the two-field volume/dilaton analysis in order to rule out certain scenarios like those proposed in \cite{Hertzberg:2007wc,Flauger:2008ad}. 
\item
Step-2: Subsequently, in a more general analysis, one also includes the dynamics of the complex-structure moduli in order to further test the de-Sitter solutions obtained by evading the constraints arising from the two-field (volume/dilaton) analysis of Step-1.
\end{itemize}
 Using the derivatives for generic scalar potential as given in Eq. (\ref{eq:gpot1}) along with (\ref{eq:typeIIA-genpot3}) w.r.t. the volume modulus $\rho$ and the dilaton $D$, we find the following extremization conditions,
\bea
\label{eq:gderV1}
& \frac{\partial V}{\partial D}= 0  \quad \Longrightarrow  \quad & 4 \, V_{f_6} + 4 \, V_{f_4} + 4 \, V_{f_2} + 4 \, V_{f_0} + 2 \, V_{h} + 2 \, V_{\omega} + 3\,V_{\rm loc} = 0, \nonumber\\
& \rho \frac{\partial V}{\partial \rho} = 0 \quad  \Longrightarrow  \quad & 3 \, V_{f_6} + \, V_{f_4} -\, V_{f_2} -3 \, V_{f_0} + 3 \, V_{h} +\, V_{\omega} = 0,
\eea
Moreover, the volume-dilaton sector of the Hessian matrix can be given as below,
\bea
\label{eq:gVij-canonical}
& & {\cal H}_{11} = 20 V_{f_6} + 20 V_{f_4} + 20 V_{f_2} +20 V_{f_0} + 6 V_{h} + 6 V_{\omega} + 12 V_{\rm loc}, \nonumber\\
& & {\cal H}_{12} = 12 V_{f_6} + 4 V_{f_4} - 4 V_{f_2} - 12 V_{f_0} + 6 V_{h} + 2 V_{\omega} = {\cal H}_{21} ,\\
& & {\cal H}_{22} = 9 V_{f_6} + V_{f_4} + V_{f_2} +9 V_{f_0} + 9 V_{h} +  V_{\omega}.\nonumber
\eea
In a nutshell, in order to investigate the possibility of de-Sitter realization, we need to check if there is a simultaneous solution to the following set of conditions (e.g. see \cite{Shiu:2011zt}),
\bea
\label{eq:no-go1}
& & 4 \, V_{f_6} + 4 \, V_{f_4} + 4 \, V_{f_2} + 4 \, V_{f_0} + 2 \, V_{h} + 2 \, V_{\omega} + 3\,V_{\rm loc} = 0, \\
& & 3 \, V_{f_6} + \, V_{f_4} -\, V_{f_2} -3 \, V_{f_0} + 3 \, V_{h} +\, V_{\omega} = 0, \nonumber\\
& & V \equiv V_{f_6} + V_{f_4} + V_{f_2} + V_{f_0} + V_{h} + V_{\omega} + V_{\rm loc} > 0,\nonumber\\
& & Tr[{\cal H}] > 0, \qquad \frac{(Tr[{\cal H}])^2}{4} \geq Det[{\cal H}] > 0. \nonumber
\eea
Using Eqn.~(\ref{eq:no-go1}), one can check that several possible scenarios simply do not allow any dS solutions at all. For example, if one does not include the geometric flux and/or the Romans mass term, it is straight forward to check that the conditions in Eqn.~(\ref{eq:no-go1}) are never satisfied, which is something observed in \cite{Hertzberg:2007wc,Flauger:2008ad}. Therefore, one has to include non-zero geometric flux as well as  Romans mass term as a requirement for evading the simplest kind of dS no-go constraint. Subsequently, there can be many solutions in support of finding the de-Sitter solution, however note that this analysis considers only the volume-dilaton sector and does not include all moduli, and hence that can have possibility to further rule out the solutions allowed at this stage. However, in case there are some no-go against finding de-Sitter in this sector itself, then there is no need to include all the moduli together, and the result remains conclusive. Now we present a couple of scenarios in which dS no-go constraints can be evaded.

\subsubsection*{Scenario 1:}
\bea
\label{eq:scenario1}
& & \hskip-0.75cm V_{f_2} = 0, \quad V_{f_6} = 0, \quad V_{f_4} > 0, \quad 10 V_{f_0} + 2 V_{f_4} + 3 V_{loc} = 4 V_h, \quad 3 V_{f_0} = V_{f_4} + 3 V_h + V_\omega, \nonumber\\
& & \hskip-0.75cm V_h \geq 0, \quad 24 V_{f_4} + 9 V_h + \sqrt{516 V_{f_4}^2 + 972 V_{f_4} V_h + 441 V_h^2} > 30 V_{f_0}, \quad V_{f_0} > V_{f_4} + V_h. 
\eea

\subsubsection*{Scenario 2:}
\bea
\label{eq:scenario2}
& & \hskip-0.75cm V_{f_2} = 0, \quad V_{f_4} = 0, \quad 24 V_{f_0} + 9 V_{loc} + 2 V_\omega = 6 V_h,   \\
& & \hskip-0.75cm V_h \geq 0, \quad V_\omega > 0,\quad 3 V_{f_0} < 3 V_h + 2 V_\omega, \quad 3 V_{f_0} = 3 V_{f_6} + 3 V_h + V_\omega, \nonumber\\
& & \hskip-0.75cm 27 V_h + 19 V_\omega + \sqrt{2025 V_h^2 + 1458 V_h V_\omega + 265 V_w^2} < 72 V_{f_0}. \nonumber
\eea

\subsubsection*{Scenario 3:}
\bea
\label{eq:scenario3}
& & \hskip-0.75cm {V_{f_6}}=0, \quad {V_{f_4}}=0, \quad \frac{7 V_{\omega}}{11} < {V_{f_2}}<{V_{\omega}}, \quad {V_h}>\frac{4 {V_{f_2}}^2-9 {V_{\omega}} {V_{f_2}}+5 {V_{\omega}}^2}{33 {V_{f_2}}-21 {V_{\omega}}}, \\
& & \hskip-0.75cm V_{\omega}  > 0, \quad V_{loc} = - \frac{8 V_{f_2} + 18 V_h + 10 V_\omega}{9}, \quad  {V_{f_0}}=\frac{3 {V_h}+{V_\omega} -{V_{f_2}}}{3}.\nonumber
\eea
In the next section, we will present some concrete models where we will present benchmark examples with specific choice of fluxes such that the necessary conditions of some of these scenarios can be met along with realizing the basic EFT requirements such as large volume, large complex structure and weak coupling values at the minimum.


\subsection{Investigating the possible flux scalings}
In this subsection we will explore the possible flux scaling which could be relevant in order to have some estimates of moduli VEVs in the process of moduli stabilization in some generic fashion. For this we need the following axionic flux combinations which are needed to express the scalar potential,
\bea
\label{eq:axionic-f-fluxes}
& & \hskip-0.3cm {\rm f}_0  = e_0 + \, {\rm b}^a\, e_a + \frac{1}{2} \, \kappa_{abc} \, {\rm b}^a\, {\rm b}^b \,m^c + \frac{1}{6}\, \kappa_{abc}\,  {\rm b}^a \, {\rm b}^b\, {\rm b}^c \, m_0 \\
& & \hskip0.3cm - \, \xi^{0} \, ({\rm H}_0 + {\rm b}^a \, {w}_{a0}) - \, \xi^k \, ({\rm H}_k + {\rm b}^a \, {w}_{ak}) - {\xi}_\lambda \, ({\rm H}^\lambda + {\rm b}^a \, {w}_{a}{}^\lambda) \,, \nonumber\\
& & \hskip-0.3cm {\rm f}_a = e_a + \, \kappa_{abc} \,  {\rm b}^b \,m^c + \frac{1}{2}\, \kappa_{abc}\,  {\rm b}^b\, {\rm b}^c \, m_0 - \, \xi^{0} \, {w}_{a0} - \, \xi^k \, {w}_{ak} - {\xi}_\lambda \, {w}_a{}^\lambda\,, \nonumber\\
& & \hskip-0.3cm {\rm f}^a = m^a + m_0\,  {\rm b}^a \,, \nonumber\\
& & \hskip-0.3cm {\rm f}^0 = m_0\,, \nonumber
\eea
Here $b^a$ counts the NS-NS $B_2$ axionic moduli and the $(\xi^{0}, \xi^{k})$ arise from the three-form potential $C_3$. In addition, $(e_0, e_a, m^a, m_0)$ denote to the flux quanta corresponding to the RR fluxes $(F_6, F_4, F_2, F_0)$ respectively while ${\rm H}$ and $w$ flux components corresponds to the NS-NS three-form flux $H_3$ and the geometric flux.
\subsubsection*{Scenario 1:}
As mentioned in Eq. (\ref{eq:scenario1}), demanding $V_{f_2} \propto {\rm f}^a \,g_{ab} \, {\rm f}^b = 0$ and $V_{f_6} \propto {\rm f}_0^2= 0$ simultaneously, where $g_{ab}$ is the moduli space metric, we get,
\bea
& & e_0 = \, \frac{m^a\, e_a}{m_0} - \frac{1}{3} \, \kappa_{abc} \, \frac{m^a\, m^b\, m^c}{m_0^2} +\, \xi^{0} \, {\rm H}_0 + \, \xi^k \, {\rm H}_k + {\xi}_\lambda \, {\rm H}^\lambda\\
& & \hskip0.3cm - \, \frac{m^a}{m_0} \,\left(\xi^{0} \, {w}_{a0} + \, \xi^k \, {w}_{ak} + {\xi}_\lambda \, {w}_{a}{}^\lambda \right), \quad b^a  = - \frac{m^a}{m_0}. \nonumber
\eea
This shows that $(h^{1,1}_- + 1)$ number of axions are fixed. Noting the fact there one can always make a rotation of the flux orbits \cite{Marchesano:2020uqz} such that all the axionic dependences are absorbed in the new flux combinations like those clubbed in Eqn.~(\ref{eq:axionic-f-fluxes}). With this observation one can satisfy both the demands $V_{f_2} = 0$ and $V_{f_6} = 0$ along with satisfying $m^a = 0$ and $e_0 = 0$ subject to imposing the following conditions on the axions,
\bea
& & b^a = 0, \qquad \xi^{0} \, {\rm H}_0 +\, \xi^k \, {\rm H}_k  +\, {\xi}_\lambda \, {\rm H}^\lambda = 0,
\eea
which results in a stronger constraint on the axions. Moreover, let us consider the flux scalings relevant for the saxionic moduli by looking into the scalar potential pieces  given in Eq. (\ref{eq:typeIIA-genpot3}). Assuming that $V_{f_2} = 0 = V_{f_6}$ and all remaining terms to be non-zero and comparable to one another at the extremum, we anticipate the following flux scaling for the corresponding moduli VEVs,
\bea
\label{eq:flux-scalings-1}
& & \rho \sim \sqrt{\frac{A_3}{A_1}} \sim \sqrt{\frac{A_5}{A_8}}, \qquad e^{D} \sim \frac{1}{\rho^3\, \sigma^{3/2}}\, \sqrt{\frac{A_5}{A_1}} \sim \frac{1}{\sigma^{3/2}}\, \frac{A_8}{A_3}, \\
&& \sigma^2 \sim \frac{A_5}{A_6} \sim \frac{A_6}{A_7} \sim \frac{A_8}{A_9} \sim \frac{A_9}{A_{10}} \sim \frac{A_{10}}{A_{11}}\sim \frac{A_{12}}{A_{13}}. \nonumber
\eea

\subsubsection*{Scenario 2:}
As mentioned in Eq. (\ref{eq:scenario2}), demanding $V_{f_2} \propto {\rm f}^a \,g_{ab}\, {\rm f}^b = 0$ and $V_{f_4} \propto {\rm f}_a \,g^{ab} \, {\rm f}_b= 0$ simultaneously we get,
\bea
& & e_a = \frac{1}{2}\, \frac{\kappa_{abc} \, m^b \,m^c}{m_0} + \, \xi^{0} \, {w}_{a0} + \, \xi^k \, {w}_{ak} + {\xi}_\lambda \, {w}_a{}^\lambda, \quad b^a  = - \frac{m^a}{m_0}. 
\eea
In fact one can satisfy $V_{f_2} = 0$ and $V_{f_4} = 0$ along with satisfying $m^a = 0$ and $e_a = 0$ subject to imposing the following conditions on the axions,
\bea
& & b^a = 0, \qquad \xi^{0} \, {w}_{a0} + \, \xi^k \, {w}_{ak} + {\xi}_\lambda \, {w}_a{}^\lambda = 0,
\eea
which results in a stronger constraint on the axions. Similar to the previous scenario, the flux scalings relevant for the saxionic moduli can be estimated by looking into the scalar potential pieces given in Eq. (\ref{eq:typeIIA-genpot3}). Assuming that $V_{f_2} = 0 = V_{f_4}$ and all remaining terms to be non-zero and comparable to one another at the extremum, we anticipate the following flux scaling for the corresponding moduli VEVs,
\bea
\label{eq:flux-scalings-2}
& & \rho \sim \left(\frac{A_4}{A_1}\right)^{1/6} \sim \sqrt{\frac{A_5}{A_8}}, \qquad e^{D} \sim \frac{1}{\rho^3\, \sigma^{3/2}}\, \sqrt{\frac{A_5}{A_1}}, \\
&& \sigma^2 \sim \frac{A_5}{A_6} \sim \frac{A_6}{A_7} \sim \frac{A_8}{A_9} \sim \frac{A_9}{A_{10}} \sim \frac{A_{10}}{A_{11}}\sim \frac{A_{12}}{A_{13}}. \nonumber
\eea

\subsubsection*{Scenario 3:}
As mentioned in Eq. (\ref{eq:scenario3}), demanding $V_{f_6} \propto {\rm f}_0^2 = 0$ and $V_{f_4} \propto {\rm f}_a \,g^{ab} \, {\rm f}_b = 0$ simultaneously we get,
\bea
& & e_a + \, \kappa_{abc} \,  {\rm b}^b \,m^c + \frac{1}{2}\, \kappa_{abc}\,  {\rm b}^b\, {\rm b}^c \, m_0 = \, \xi^{0} \, {w}_{a0} +\, \xi^k \, {w}_{ak} + {\xi}_\lambda \, {w}_a{}^\lambda, \\
& & e_0 = \frac{1}{2} \, \kappa_{abc} \, {\rm b}^a\, {\rm b}^b \,m^c + \frac{1}{3}\, \kappa_{abc}\,  {\rm b}^a \, {\rm b}^b\, {\rm b}^c \, m_0\, +\, \xi^{0} \, {\rm H}_0  + \, \xi^k \, {\rm H}_k + {\xi}_\lambda \, {\rm H}^\lambda \,. \nonumber
\eea
In fact one can satisfy $V_{f_6} = 0$ and $V_{f_4} = 0$ along with satisfying $e_0 = 0$ and $e_a = 0$ subject to imposing the following conditions on the axions,
\bea
& & \xi^{0} \, {w}_{a0} +\, \xi^k \, {w}_{ak} + {\xi}_\lambda \, {w}_a{}^\lambda = \kappa_{abc} \,  {\rm b}^b \,m^c + \frac{1}{2}\, \kappa_{abc}\,  {\rm b}^b\, {\rm b}^c \, m_0, \\
& & \xi^{0} \, {\rm H}_0  + \, \xi^k \, {\rm H}_k + {\xi}_\lambda \, {\rm H}^\lambda = - \frac{1}{2} \, \kappa_{abc} \, {\rm b}^a\, {\rm b}^b \,m^c - \frac{1}{3}\, \kappa_{abc}\,  {\rm b}^a \, {\rm b}^b\, {\rm b}^c \, m_0\,. \nonumber
\eea
Similar to the previous scenario, the flux scalings relevant for the saxionic moduli can be estimated by looking into the scalar potential pieces given in Eq. (\ref{eq:typeIIA-genpot3}). Assuming that $V_{f_4} = 0 = V_{f_6}$ and all remaining terms to be non-zero and comparable to one another at the extremum, we anticipate the following flux scaling for the corresponding moduli VEVs,
\bea
\label{eq:flux-scalings-3}
& & \rho \sim \sqrt{\frac{A_2}{A_1}} \sim \sqrt{\frac{A_5}{A_8}}, \qquad e^{D} \sim \frac{1}{\rho^3\, \sigma^{3/2}}\, \sqrt{\frac{A_5}{A_1}}, \\
&& \sigma^2 \sim \frac{A_5}{A_6} \sim \frac{A_6}{A_7} \sim \frac{A_8}{A_9} \sim \frac{A_9}{A_{10}} \sim \frac{A_{10}}{A_{11}}\sim \frac{A_{12}}{A_{13}}. \nonumber
\eea
We will discuss the relevance of  these flux scalings in more explicit settings in the upcoming section. 


\subsection{Constructing a concrete type IIA orientifold model}

In \cite{Shukla:2019dqd} it has been shown that the type IIA geometric flux models and a certain class of the type IIB non-geometric models (with special solutions of Bianchi identities) are $T$-dual  to each other in the sense that there are certain transformations on the moduli/fluxes which exchange the scalar potential on the one theory with the same on the other. Now, we consider a simple type IIA construction which will be used to investigate the possibility of realizing tachyon-free AdS, Minkowskian and dS solutions, and their type IIB analogue can be generically read-off from the dictionary presented in \cite{Shukla:2019dqd}. 

In order to make some simplification for the explicitness of the computations, we consider the following topological quantities for the compactifying threefold \footnote{Given the fact that splitting of the $h^{2,1}$ index in $\{\hat{k}, \lambda\} \equiv \{0, k, \lambda\}$ is such that $k + \lambda = h^{2,1}$, and so for our particular model, we take $k =0$ and $\lambda =1$. This means that the complex variables ${\rm N}^k$, which are like the so-called odd-moduli $G^a$ in the dual Type IIB picture, will be absent along with the flux components involving the $k$-indices.},
\bea
& & h^{1,1}_- = 1, \quad \quad h^{1,1}_+ = 1, \quad \quad h^{2,1} =1.
\eea
Such an orientifold leads to the following fluxes/moduli in the 4D type IIA model, which has six real moduli and 10 flux parameters to begin with. These are collected as below,
\bea
& {\rm F \, \, term \, \, fluxes}: & \quad e_0, \quad e_1, \quad m^1, \quad m_0, \quad {\rm H}_0, \quad {\rm H}^1, \quad {w}_{10}, \quad {w}_{1}{}^1, \\
& {\rm D \, \, term \, \, fluxes}: & \quad \hat{w}_1{}^{0}, \quad \hat{w}_{1 1}, \nonumber\\
& {\rm Moduli}: & \quad \rho, \quad \sigma, \quad D, \quad {\rm b}^1, \quad \xi^{1}, \quad \xi^{0}. \nonumber
\eea
We denote D-term fluxes with a hat in order to distinguish them from the geometric fluxes appearing in the F-term contributions. Subsequently, the total scalar potential arising from F- and D-terms can be given by Eqn.~(\ref{eq:typeIIA-genpot3}) with the following simpler form of the $A_i$ coefficients,
 \bea
\label{eq:coeff-Ais}
& & A_1 = \frac14\, {({\rm f}^0)}^2, \quad A_2 = \frac34\, ({\rm f}^1)^2 \,, \quad A_3 = \frac{1}{12}\,  {\rm f}_1^2 \, , \quad A_4 = \frac14\,{\rm f}_0^2, \\
& & A_5 = \frac14\,({\rm H}_0 + {\rm b}^1 \, {w}_{10})^2, \quad A_6 =0, \quad A_7 = \frac34\, ({\rm H}^1 + {\rm b}^1 \, {w}_{1}{}^1)^2 \, , \quad A_8 = \frac{1}{12}\,w_{10}^2, \nonumber\\
& & A_9 = w_{10} \,w_1{}^1 + \frac{1}{4\,\hat\kappa_{111}} \,\hat{w}_{11}^2\,, \quad  A_{10}= - \frac34\, (w_1{}^1)^2 + \frac{1}{2\,\hat\kappa_{111}} \,\hat{w}_{11}\,\hat{w}_1{}^{0}\,, \quad A_{11} = \frac{1}{4\,\hat\kappa_{111}} \, (\hat{w}_1{}^{0})^2, \nonumber\\
& & A_{12} = \frac12 \,\left(m_0 \, {\rm H}_0 - m^1\, w_{10}\right), \quad A_{13} = -\frac32 \, \left(m_0\,{\rm H}^1 - m^1\, w_1{}^1 \right). \nonumber
\eea
where we have used the triple intersection number $\kappa_{111} = 6$ and the axionic flux combinations ${\rm f}_0$, ${\rm f}_1$ etc. are defined as below,
\bea
\label{eq:axionic-flux-nogo2}
& & \hskip-0.3cm {\rm f}_0  = e_0 + \, {\rm b}^1\, e_1 + 3 \, ({\rm b}^1)^2 \,m^1 +  ({\rm b}^1)^3 \, m_0 - \, \xi^{0} \, ({\rm H}_0 + {\rm b}^1 \, {w}_{10}) - {\xi}_1 \, ({\rm H}^1 + {\rm b}^1 \, {w}_{1}{}^1) \,, \nonumber\\
& & \hskip-0.3cm {\rm f}_1 = e_1 + \, 6 \, {\rm b}^1 \,m^1 + 3\,  ({\rm b}^1)^2 \, m_0 - \, \xi^{0} \, {w}_{10} - {\xi}_1 \, {w}_1{}^1\,, \\
& & \hskip-0.3cm {\rm f}^1 = m^1 + \,  {\rm b}^1\,m_0 \,, \quad  {\rm f}^0 = m_0\,, \nonumber
\eea
Here we note that all the coefficients except $A_9, A_{10}, A_{12}$ and $A_{13}$ are positive semidefinite, which is something which we have pointed out earlier for the generic case as well. In other words, apart from the local/tadpole piece in the scalar potential with a dilaton factor $e^{3D}$ encoded in the coefficients $A_{12}$ and $A_{13}$, all the other pieces are positive semidefinite except for the two pieces involving the geometric flux $w_1{}^1$ as seen from the coefficients $A_9$ and $A_{10}$. Therefore, for the purpose of hunting for the de-Sitter solutions, it would be a good idea to set this geometric flux component to zero, i.e. $w_1{}^1 = 0$. However this will subsequently have an impact on the following two Bianchi identities, 
\bea
\label{eq:BIs-simp}
& & \hskip-1cm {\rm H}^{1} \, \hat{w}_{11} = {\rm H}_{0} \, \hat{w}_1{}^{0}, \qquad \qquad w_1{}^1 \, \hat{w}_{11} = w_{10} \, \hat{w}_1{}^{0}.
\eea
Setting $w_1{}^1 = 0$ will demand either to set $w_{10} = 0$ or $\hat{w}_1{}^{0} = 0$. However, given that we do not want to switch-off any of the (positive semidefinite) D-term piece in the scalar potential which could possibly be useful to make the dS uplift, we take $w_{10} = 0$. This means that we are switching-off all the geometric flux components in the F-term contributions, and the only geometric flux effects are those arising from the D-term contributions.


\section{Analyzing the flux vacua}
\label{sec_AdS-dS-vacua}
It has been observed in \cite{Shiu:2011zt} that having a non-zero Romans mass $m_0$ with the presence of any one of the ${\rm F}_2, {\rm F}_4$ or ${\rm F}_6$ fluxes can evade the simplest version of the dS no-go conditions obtained in the volume-dilaton analysis. This can be also seen from the Scenario-1 which we presented in the previous section. Now we take the flux choice for the RR flux such that 
\bea
\label{eq: RR-IIA}
& & e_0 = 0, \qquad \qquad m^1 = 0\,.
\eea
The type IIA scalar potential which we have at this stage can be expressed in the following simple form,
\bea
\label{eq:V-case-II}
& & \hskip-0.3cm V =  \frac{e^{4D}}{4\,\rho^3} \biggl[({\rm f}_0)^2 +  \frac{\rho^2}{3} \, ({\rm f}_1)^2 \, + 3\, \rho^4\, ({\rm f}^1)^2  + \rho^6\, ({\rm f}^0)^2\biggr] \\
& & + \frac{e^{2D}}{4\,\rho^3\, \sigma^3}\biggl[{\rm H}_0^2 + 3\, \sigma^4 ({\rm H}^1)^2 + \, \frac{\rho^2\, \sigma^2\, \hat{w}_{1 1}^2}{\hat\kappa_{111}} \, \left(1 + \frac{\sigma^2 \, {\rm H}^1}{{\rm H}_0}  \right)^2 \biggr] + \frac{e^{3D}}{2\, \sigma^{3/2}} \left[m_0 \left({\rm H}_0 - 3\,\sigma^2 \,  {\rm H}^1 \right) \right],\nonumber
\eea
where we have used the first Bianchi identity of Eqn.~(\ref{eq:BIs-simp}) along with the following simplified axionic flux combinations,
\bea
\label{eq:axionic-flux-nogo2-1}
& & \hskip-0.3cm {\rm f}_0  = {\rm b}^1\, e_1 + \,  ({\rm b}^1)^3 \, m_0 - \, \xi^{0} \, {\rm H}_0 - {\xi}_1 \, {\rm H}^1, \quad {\rm f}_1 = e_1 + 3\,  {\rm b}^1\, {\rm b}^1 \, m_0\,, \quad  {\rm f}^1 = \,  {\rm b}^1\,m_0 \,, \quad  {\rm f}^0 = m_0. 
\eea
Using the scalar potential in Eqns.~(\ref{eq:V-case-II})-(\ref{eq:axionic-flux-nogo2-1}), it turns out that the extremization of three axions $\xi^0$, $\xi_1$ and ${\rm b}^1$ results in satisfying the following two conditions, 
\bea
& & {\rm f}_0 = {\rm f}^1 = 0,
\eea
where we keep the Romans mass parameter ${\rm f}^0 = m_0 \neq 0$ as that is necessary to avoid the well-known de-Sitter no-go theorems for geometric type IIA scenarios \cite{Hertzberg:2007ke,Flauger:2008ad}. Using Eqn.~(\ref{eq:axionic-flux-nogo2-1}), we find that this choice of RR-flux helps in setting axions ${\rm b}^1$ to zero and subsequently, only two (out of three) axionic moduli get stabilized as below,
\bea
& & \langle {\rm b}^1 \rangle = 0, \qquad \quad \langle \xi^0 \rangle\, {\rm H}_0 + \langle \xi_1 \rangle\, {\rm H}^{1} = 0\,.
\eea
We note that the process of axion minimization has also set $A_1 = \frac{1}{4}\, m_0^2, A_2 = 0 = A_4$ and $A_3 = \frac{1}{12}\, e_1^2$. Now one can have some estimates about the flux scalings corresponding to the simplified scalar potential in Eqn.~(\ref{eq:V-case-II}) which turns out to be given as below,
\bea
\label{eq:scaling1}
& & e^{-D} \sim \frac{(e_1)^\frac32}{(m_0)^\frac12\,({\rm H}_0)^\frac14\, ({\rm H}^1)^\frac34}, \quad \rho \sim \sqrt{\frac{e_1}{m_0}}, \quad \sigma \sim \sqrt{\frac{{\rm H}_0}{{\rm H}^1}},
\eea
which shows that the string coupling $g_s = e^{\langle \varphi \rangle}$ obtained from the VEV of the ten dimensional dilaton ($\varphi$) should scale as follows,
\bea
\label{eq:scaling2}
& & e^{-\varphi} = \frac{e^{-D}}{\sqrt{\cal V}} \sim \frac{(e_1)^\frac34\, (m_0)^\frac14}{({\rm H}_0)^\frac14\,({\rm H}^1)^\frac34}.
\eea
Note that similar flux scalings for saxion VEVs have been observed for the rigid toroidal orientifold of ${\mathbb T}^6/({\mathbb Z}_3 \times {\mathbb Z}_3)$ in \cite{DeWolfe:2005uu}, however we also have complex-structure moduli in our setup.

\subsection{AdS vacua: without $D$-terms}

Now, let us have the exact conditions for moduli stabilization in the absence of $D$-term effects, with the aim to engineer the fluxes so that we could look for the possibility of de-Sitter solutions by adding such effect as a second step. 

In the absence of the $D$-term contribution, i.e. for $\hat{w}_{11} = 0 = \hat{w}_1{}^0$, the saxionic extremization conditions can be satisfied for the following two sets of AdS solutions:
\bea
& & \hskip-1.5cm {\bf AdS1:} \qquad e^{-\langle D \rangle} = -\frac{2 \, m_0}{5 \,{\rm H}_0} \, {\langle \rho\rangle^3}\,{\langle \sigma \rangle^\frac32}, \qquad \langle \sigma \rangle^2 = - \frac{{\rm H}_0}{{\rm H}_1}, \qquad \langle \rho \rangle^2 = \pm \frac{5 \,e_1}{9 \,m_0}, \\
& & \hskip0.5cm \langle V \rangle = -\frac{3\, e^{4\langle D \rangle}\, \langle \rho \rangle^3\, m_0^2}{25} = -\frac{3\, e^{2\langle D \rangle}\, {\rm H}_0^2}{4\,\langle \rho \rangle^3\, \langle \sigma \rangle^3 }; \nonumber\\
& &  \hskip-1.5cm {\bf AdS2:} \qquad e^{-\langle D \rangle} = -\frac{4 \,m_0}{5 \,{\rm H}_0} \, {\langle \rho\rangle^3}\,{\langle \sigma \rangle^\frac32}\,, \qquad \langle \sigma \rangle^2 = - \frac{{\rm H}_0}{4\,{\rm H}^1}, \qquad \langle \rho \rangle^2 = \pm \frac{5 \,e_1}{3\,\sqrt{6}\, m_0}, \\
& & \hskip0.5cm \langle V \rangle = -\frac{2\, e^{4\langle D \rangle}\, \langle \rho \rangle^3\, m_0^2}{25} = -\frac{\, e^{2\langle D \rangle}\, {\rm H}_0^2}{8\,\langle \rho \rangle^3\, \langle \sigma \rangle^3}. \nonumber
\eea
The Hessian analysis further shows that ${\bf AdS1}$ corresponds to a tachyonic solution while ${\bf AdS2}$ is a minimum. Given that saxionic VEVs have to be positive, we need to choose $m_0$ and ${\rm H}_0$ of opposite sign, and subsequently there can be two possibilities for choosing the sign of the fluxes,
\bea
& (i). & \quad m_0 > 0, \quad e_1 < 0, \quad {\rm H}_0 < 0, \quad {\rm H}^1 > 0, \\
& (ii). & \quad m_0 < 0, \quad e_1 > 0, \quad {\rm H}_0 > 0, \quad {\rm H}^1 < 0. \nonumber
\eea
Note that the tadpole contributions in both the cases are $N_{\rm flux} < 0$ which is consistent with the standard literature of type IIA models without geometric flux (e.g. see \cite{DeWolfe:2005uu, Narayan:2010em}). Without loss of any generality, let us take the first choice. Subsequently, the two AdS solutions can be equivalently expressed entirely in terms of fluxes as below,
\bea
\label{eq:twoAdS-IIA}
& & \hskip-0.1cm {\bf AdS1:} \quad e^{-\langle D \rangle} = \frac{2 \sqrt{5} \, (-e_1)^\frac32}{27 \sqrt{m_0} \, (-{\rm H}_0)^\frac14\,({\rm H}^1)^\frac34\,}, \quad \langle \rho \rangle = \frac{\sqrt{5}}{3} \sqrt{-\frac{e_1}{m_0}}, \quad \langle \sigma \rangle = \sqrt{- \frac{{\rm H}_0}{{\rm H}^1}}, \qquad \\
& & \hskip1.5cm e^{-\langle\varphi \rangle} = \frac{2\,(-e_1)^\frac34\, m_0^\frac14}{3^\frac32\,5^\frac14\,(-{\rm H}_0)^\frac14\,({\rm H}^1)^\frac34},  \quad \langle V \rangle = -\, \frac{59049 \, (m_0)^\frac52\, (-{\rm H}_0)\,({\rm H}^1)^3}{400\,\sqrt{5} \, (-e_1)^\frac92}; \nonumber\\
& &  \hskip-0.1cm {\bf AdS2:} \quad e^{-\langle D \rangle} = \frac{\sqrt{5} \, (-e_1)^\frac32}{9 \times 6^\frac14\, \sqrt{m_0} \, (-{\rm H}_0)^\frac14\,({\rm H}^1)^\frac34}, \quad \langle \rho \rangle = \frac{\sqrt{5}}{2^\frac14\, 3^\frac34} \sqrt{-\frac{e_1}{m_0}}, \quad \langle \sigma \rangle = \frac{1}2{}\sqrt{- \frac{{\rm H}_0}{{\rm H}^1}}, \qquad \nonumber\\
& & \hskip1.5cm e^{-\langle\varphi\rangle} = \frac{2^\frac18\,(-e_1)^\frac34\, m_0^\frac14}{3^\frac98\,5^\frac14\,(-{\rm H}_0)^\frac14\,({\rm H}^1)^\frac34}, \quad \langle V \rangle = -\, \frac{1458\times 2^\frac14\, 3^\frac34 \, m_0^\frac52\, (-{\rm H}_0)\,({\rm H}^1)^3}{25\,\sqrt{5} \, (-e_1)^\frac92}. \nonumber
\eea
These analytic results are extremely useful in the many ways. First let us observe that the exact AdS solutions in Eq. (\ref{eq:twoAdS-IIA}) confirm the previous estimates about the flux scalings for the saxionic VEVs as given in Eqs. (\ref{eq:scaling1})-(\ref{eq:scaling2}). Also such flux scaling of the saxionic VEVs can help in determining the flux regions where the solutions can be physically acceptable and trustworthy. In this regard, we make the following points:
\begin{itemize}
\item{We observe that choosing suitably large values for the $|e_1|$ flux can be used to realize larger VEVs for volume modulus $\rho$ along with weaker values for the string coupling $g_s = e^{\langle \varphi \rangle}$. Given that there are no non-geometric fluxes present in the model, the $e_1$ flux is not restricted by the tadpole condition as well.} 
\item{Also, we observe that the $m_0$ flux should be taken to a minimum consistent value in order to realize large VEVs for volume modulus $\rho$ as well as the having weak coupling, and therefore we prefer to set $m_0 = 1$ in our numerical analysis.}
\item{It is obvious that the ${\rm H}^1$ flux should be taken to minimum values in order to have larger VEV for the complex structure modulus $\sigma$ along with weak string coupling, and therefore we prefer to set ${\rm H}^1 = 1$ in our numerical analysis. 
}
\item{Although larger values for ${\rm H}_0$ could also help for realizing large VEVs for the complex-structure modulus $(\sigma)$, however this can happen only at the cost of enhancing the string coupling by a reduction of the dilaton factor $e^{-\varphi}$. Thus we observe that large complex structure and weak coupling requirements are apparetnly contradictory to each other with respect to the choice of the ${\rm H}_0$ flux, and one has to find a viable balance to keep both in the physically valid regime.}
\item{However, choosing larger values for $|e_1|$ flux can help in having weak-coupling and large volume realizations, and also leaves a scope for large complex structure realization by keeping the factor $|e_1|/|{\rm H}_0|$ large while having $|{\rm H}_0|$ large as well.}
\item{With these scaling arguments, we find that the moduli dynamics can be mainly controlled by two fluxes $|e_1|$ and $|{\rm H}_0|$. However, demanding $|e_1|/{\rm H}_0$ and ${\rm H}_0$ to take large values has to be balanced by the need of large tadpole charge compensation, and so one would not prefer to take too large values for $|{\rm H}_0|$ and should be satisfied with those which are just enough to ensure $\langle \sigma \rangle > 1$.}
\item{A couple of benchmark numerical samplings are presented in Table \ref{tab_samplings1-IIA} and \ref{tab_samplings2-IIA}.}
\end{itemize}

\vskip-0.2cm
\noindent
\begin{center}
\renewcommand\arraystretch{1}
\begin{longtable}{|c||c|c||c|c|c||c|c|c||} 
\caption{Numerical samplings for tachyonic and stable AdS solutions where other fluxes are set as: $e_0 = 0, \, m^1 =0, \, m_0 = 1, \, {\rm H}^1 =1$ and $\hat{w}_{11} =0$, leading to (two of the three) axions being stabilized as $\langle b^1 \rangle= 0$ and $\langle \xi^0 \rangle \, {\rm H}_0 + \langle \xi_1 \rangle\, {\rm H}^{1} = 0$.}\\
\hline
Sample & $e_1$  & ${\rm H}_0$ & $\langle\rho\rangle$ & $\langle\sigma\rangle$ & $\langle e^{-D}\rangle$ & $\langle{\cal V}\rangle$ & $\langle e^{-\varphi}\rangle$ &  Vacua type\\
\hhline{|=|=|=|=|=|=|=|=|=|}
\endhead
\label{tab_samplings1-IIA}
\bf{S1} & -50 & -5 & 5.27046 & 2.23607 & 39.1619 & 146.402 & 3.23661 &Tachyonic AdS \\
 & & & 5.83273 & 1.11803 & 37.5333 & 198.433 & 2.66446 & Stable AdS  \\
\hline
 \bf{S2} & -75 & -5  & 6.45497 & 2.23607 & 71.945 & 268.957 & 4.38691 & Tachyonic AdS \\
 & & & 7.1436 & 1.11803 & 68.9531 & 364.545 & 3.61142 & Stable AdS \\
\hline
 \bf{S3} & -100 & -5 & 7.45356 & 2.23607 & 110.767 & 414.087 & 5.44331 & Tachyonic AdS \\
 & & & 8.24872 & 1.11803 & 106.16 & 561.254 & 4.48108 & Stable AdS \\
\hline
  \bf{S4} & -100 & -50  & 7.45356 & 7.07107 & 62.2886 & 414.087 & 3.061 & Tachyonic AdS \\
 & & & 8.24872 & 3.53553 & 59.6983 & 561.254 & 2.51989 & Stable AdS \\
\hline
\end{longtable}
\end{center}

\vskip-1.5cm
\noindent
\begin{center}
\renewcommand\arraystretch{1}
\begin{longtable}{|c||c|c||} 
\caption{Hessian Eigenvalues and the scalar potential VEVs corresponding to the AdS solutions for the flux samplings presented in Table \ref{tab_samplings1-IIA}. This shows that the first AdS solution is tachyonic while the second AdS solution is tachyon free, but having a flat axionic combination.}\\
\hline
Sample & $\langle V_0 \rangle \cdot 10^6$ & Eigenvalues of $\langle V_{ij} \rangle \cdot 10^6$ \\
\hhline{|=|=|=|}
\endhead
\label{tab_samplings2-IIA}
\bf{S1} & -7.46917 & $\{8.88289 , 3.00399 , -1.49383 , 0.0619177 , -0.00844809 , 0 \}$ \\
& -7.99900 & $\{50.7392 , 3.29533 , 1.05382 , 0.0647094 , 0.000892266 , 0 \}$ \\
\hline
\bf{S2} & -1.20465 & $\{0.954538 , 0.32024 , -0.240931 , 0.00296067 , -0.000407189 , 0 \}$ \\
& -1.29011 & $\{7.57355 , 0.352403 , 0.122203 , 0.00309812 , 0.0000428714 , 0 \}$ \\
\hline
\bf{S3} & -0.330094 & $\{0.196128 , -0.0660188 , 0.0656144 , 0.000342326 , -0.0000472137 , 0 \}$ \\
& -0.353509 & $\{1.99277 , 0.0722849 , 0.0261409 , 0.000358322 , 4.9654\times10^{-6} , 0 \}$ \\
\hline
 \bf{S4} & -3.30094 & $\{ 1.96242 , 0.887765 , -0.0660188 , -0.0335668 , 0.010819, 0\}$ \\
 & -3.53509 & $\{ 4.32243 , 0.892626 , 0.121494 , 0.0112401 , 0.00386789 , 0\}$ \\
\hline
\end{longtable}
\end{center}


\subsection{dS vacua: with $D$-terms}
In this section we investigate the effects of including $D$-term contributions which are generically positive semi-definite in nature, and hence one may expect to realize stable de-Sitter solution. Just to have some naive estimates (which may or may not be true as we will explore later on), momentarily if we simply assume that the stabilized values of the moduli/axions do not change significantly, then the $D$-term contributions to the scalar potential at the previously realized AdS minimum in Eqn.~(\ref{eq:twoAdS-IIA}) may be given as,
\bea
& & \hskip-1cm \langle V_D \rangle  = \left\langle\frac{e^{2D}}{4\,\rho\, \sigma}\frac{\hat{w}_{1 1}^2}{\hat\kappa_{111}} \, \left(1 + \frac{\sigma^2 \, {\rm H}^1}{{\rm H}_0}  \right)^2 \right\rangle= \, \frac{2187\times 3^\frac14  \, (m_0)^\frac32\,({\rm H}^1)^2\, \hat{w}_{11}^2}{80 \,\sqrt{5} \times 2^\frac14\, \hat\kappa_{111}\, (-e_1)^\frac72},
\eea 
where we have used the saxion/axion VEVs corresponding to the ${\bf AdS2}$ solution in (\ref{eq:twoAdS-IIA}) which is a tachyon free minimum in the absence of $D$-terms. This naive estimate shows that a priory there appears to be a chance of uplifting the AdS to some dS solution under the assumption that the previously stabilized values remain (almost) fixed at their respective minimum. For that we need,
\bea
& & \hskip-2.0cm  \langle V \rangle = \langle V_F \rangle + \langle V_D \rangle = -\, \frac{1458\times 2^\frac14\times 3^\frac34 \, (m_0)^\frac52\, (-{\rm H}_0)\,({\rm H}^1)^3}{25\,\sqrt{5} \, (-e_1)^\frac92} \\
& &  + \frac{2187\times 3^\frac14  \, (m_0)^\frac32\,({\rm H}^1)^2\, \hat{w}_{11}^2}{80 \,\sqrt{5} \times 2^\frac14\, \hat\kappa_{111}\, (-e_1)^\frac72}\, \geq 0, \nonumber
\eea
which can be satisfied if,
\bea
\label{eq:w^2}
& & \hat{w}_{11}^2 \geq \frac{32\sqrt{2}}{5\sqrt{3}}\, \times \frac{(m_0)\, (-{\rm H}_0) \,({\rm H}^1)}{(-e_1)}\, \hat\kappa_{111}.
\eea
Just to have some rough estimate, for the flux choices taken in the samples {\bf S1}-{\bf S4}, the requirement in Eq. (\ref{eq:w^2}) simplifies into the following form,
\bea
& & \hskip-1.5cm \hat{w}_{11}^2 \geq \lambda\,\hat\kappa_{111}, \quad {\rm where} \quad  \lambda = \biggl\{0.522558, 0.348372, 0.261279, 2.61279 \biggr\}. \nonumber
\eea
So there appears to be a hope of uplifting the previous AdS solution to some de-Sitter for some choice of geometric flux and the triple intersection number. However, recall again that in arriving at this naive estimate, we have assumed that the previous VEVs of saxions are not significantly changed after including the $D$-term, which may not turn out to be correct, given that it depends on all the three saxions $\varphi, \rho$ and $\sigma$ and every piece in the scalar potential is on the same footing, in the sense of having no hierarchy to begin with.

Now we turn to explicitly solving the extremization conditions to investigate about the possibility dS solutions. Using the flux scaling arguments suggested in Eq.~(\ref{eq:scaling1}) and Eq.~(\ref{eq:scaling2}) along with their validation for the absence of geometric flux scenarios in Eq. (\ref{eq:twoAdS-IIA}) we take the following ansatz to begin with,
\bea
\label{eq:ansatz}
& & \langle {\rm b}^1 \rangle = 0, \qquad \langle \xi_1 \rangle = -\frac{{\rm H}_0}{{\rm H}^1} \, \langle\xi_0\rangle, \qquad e^{-\langle D \rangle} = \frac{\alpha_1 \, (-e_1)^\frac32}{\sqrt{m_0} \, (-{\rm H}_0)^\frac14\,({\rm H}^1)^\frac34\,}, \\
& & \langle \rho \rangle = \alpha_2\sqrt{-\frac{e_1}{m_0}}, \qquad \langle \sigma \rangle = \alpha_3\,\sqrt{-\frac{{\rm H}_0}{{\rm H}^1}}. \nonumber
\eea
In addition, we introduce a new parameter $\alpha_4$ defined through the following flux ratio which is induced only through the $D$-term effects,
\bea
\label{eq:alpha4}
& & \alpha_4 = \frac{e_1\, \hat{w}_{11}^2}{\hat{\kappa}_{111}\,m_0 \,{\rm H}_0\, {\rm H}^1}.
\eea
By considering these ansatz, similar to the previously realized AdS solutions which correspond to the flux choice considered as $\{m_0 > 0, \, e_1 < 0, \, {\rm H}_0 < 0, \, {\rm H}^1 > 0\}$, we find that it does not solve the extremization conditions for $\{\alpha_1 > 0, \,\alpha_2 >0, \, \alpha_3 >0, \, \alpha_4  \geq 0\}$. Therefore we conclude that the previous AdS solutions cannot be lifted to Minkowskian or dS solution by adding the $D$-terms of the type we considered.

However we find that there can be new Minkowskian/dS solutions (which does not arise as an uplifted version of the previous AdS solutions) for a different type of flux choice given as $\{m_0 > 0, \, e_1 < 0, \, {\rm H}_0 > 0, \, {\rm H}^1 > 0\}$ and for exploring that possibility we consider the following ansatz,
\bea
\label{eq:ansatz-dS}
& & \langle {\rm b}^1 \rangle = 0, \quad \langle \xi_1 \rangle = -\frac{{\rm H}_0}{{\rm H}^1} \, \langle\xi_0\rangle, \quad \langle \tau \rangle = e^{-\langle D \rangle} = \frac{\alpha_1 \, (-e_1)^\frac32}{\sqrt{m_0} \, ({\rm H}_0)^\frac14\,({\rm H}^1)^\frac34\,} > 0, \\
& & \langle \rho \rangle = \alpha_2\sqrt{-\frac{e_1}{m_0}} > 0, \quad \langle \sigma \rangle = \alpha_3\,\sqrt{\frac{{\rm H}_0}{{\rm H}^1}} > 0, \quad \alpha_4 = \frac{(-e_1)\, \hat{w}_{11}^2}{\hat\kappa_{111}\,(m_0) \,({\rm H}_0)\, ({\rm H}^1)}, \quad \forall \alpha_i > 0.\nonumber
\eea
To elaborate more on it we use Eq. (\ref{eq:ansatz-dS}) to get the following expressions for the three saxion extremization conditions,
\bea
\label{eq:Vext-STU}
& &  \partial_\tau V  = -\frac{(m_0)^3\, ({\rm H}_0)^\frac54\, ({\rm H}^1)^\frac{15}{4}}{6 (-e_1)^6 \alpha_1^5\, \alpha_2^3 \alpha_3^3} \biggl[9 \alpha _1 \alpha _3^{3/2} \left(1-3 \alpha _3^2\right) \alpha _2^3+2 \left(3 \alpha _2^4+1\right) \alpha _3^3 \alpha _2^2 \\
& & \hskip1.5cm +3 \alpha _1^2 \left(\alpha _2^2 \alpha _4 \alpha_3^6+\left(2 \alpha _4 \alpha _2^2+3\right) \alpha _3^4+\alpha _2^2 \alpha _4 \alpha _3^2+1\right) \biggr], \nonumber\\
& &  \partial_\sigma V  = \frac{(m_0)^\frac52\, ({\rm H}_0)^\frac12\, ({\rm H}^1)^\frac72}{4\, (-e_1)^\frac92 \, \alpha_1^3 \, \alpha_2^3 \, \alpha_3^4} \biggl[\left(\alpha _3^2+1\right) \left(\alpha _1 \left(3 \alpha_2^2 \alpha _4 \alpha _3^4+\left(3-\alpha _2^2 \alpha _4\right) \alpha _3^2-3\right)-3 \alpha _2^3 \alpha _3^{3/2}\right) \biggr], \nonumber\\
& & \partial_\rho V  = -\frac{(m_0)^3\, ({\rm H}_0)\, ({\rm H}^1)^3}{12 \, (-e_1)^5 \, \alpha_1^4 \, \alpha_2^4 \, \alpha_3^3} \biggl[\alpha _2^2 \left(1-9 \alpha_2^4\right) \alpha _3^3+3 \alpha _1^2 \left(\alpha _2^2 \alpha _4 \alpha _3^6+\left(2 \alpha _4 \alpha _2^2+9\right) \alpha _3^4+\alpha _2^2 \alpha _4 \alpha _3^2+3\right)\biggr], \nonumber
\eea
while the scalar potential defined through Eqs.~(\ref{eq:V-case-II})-(\ref{eq:axionic-flux-nogo2-1}) takes the following form,
\bea
\label{eq:potVEV}
& & \hskip-1.5cm V = \frac{(m_0)^\frac52\, ({\rm H}_0)\, ({\rm H}^1)^3}{12 (-e_1)^\frac92 \alpha_1^4 \alpha_2^3 \alpha_3^3} \biggl[6 \alpha _1 \alpha _3^{3/2} \left(1-3 \alpha _3^2\right) \alpha _2^3+\left(3 \alpha _2^4+1\right) \alpha _3^3 \alpha _2^2 \\
& & \hskip1cm +3 \alpha _1^2 \left(\alpha_2^2 \alpha _4 \alpha _3^6+\left(2 \alpha _4 \alpha _2^2+3\right) \alpha _3^4+\alpha _2^2 \alpha _4 \alpha _3^2+1\right) \biggr]. \nonumber
\eea
Subsequently one can get a Minkowskian or dS solution by solving the extremization conditions, in addition to consistently demanding the following constraint, 
\bea
\label{eq:alpha4-Minkowskian}
& & \alpha_4 \geq -\frac{6 \alpha _1 \alpha _3^{3/2} \left(1-3 \alpha _3^2\right) \alpha _2^3+\left(3 \alpha _2^4+1\right) \alpha _3^3 \alpha _2^2+\alpha _1^2 \left(9 \alpha
   _3^4+3\right)}{3 \alpha _1^2 \alpha _2^2 \alpha _3^2 \left(\alpha _3^2+1\right){}^2},
\eea
where equality corresponds to the Minkowskian solution. Interestingly we find that, there is a unique numerical solution for the Minkowskian case which corresponds to solving four polynomial equations in four unknowns $\alpha_i$' resulting in,
\bea
\label{eq:alphai's-num}
& & \hskip-1cm \alpha_1 = 0.386877, \quad \alpha_2 = 1.03182, \quad \alpha_3 = 1.84403, \quad \alpha_4 = 0.424169.
\eea
Note that these $\alpha_i$'s can generically be some irrational numbers and a true Minkowskian solution will demand them to be take those precise values. However for numerical estimates we have presented rounded off figures for these $\alpha_i$ parameters.

\subsubsection*{Numerical samplings for Minkowskian solutions:}
For a set of flux choice with non-zero $D$-term flux $\hat\omega_{11}$, the results for various VEVs of the moduli/axions along with their Hessian Eigenvalues are mentioned in Table \ref{tab_samplings3a} and Table \ref{tab_samplings4a}. 
\noindent
\begin{table}[H]
\begin{center}
\begin{tabular}{|c||c|c||c|c|c||c|c||} 
\hline
Sample & $e_1$  & ${\rm H}_0$ & $\langle\rho\rangle$ & $\langle\sigma\rangle$ & $\langle e^{-D}\rangle$ & $\langle{\cal V}\rangle$ & $\langle e^{-\varphi}\rangle$ \\
\hline
\hline
\bf{S1} & -50 & 5 & 7.29606 & 4.12337 & 91.4714 & 388.388 & 4.64144 \\
 \bf{S2} & -75 & 5 & 8.93581 & 4.12337 & 168.044 & 713.514 & 6.29102 \\
 \bf{S3} & -100 & 5 & 10.3182 & 4.12337 & 258.72 & 1098.53 & 7.80593 \\
 \bf{S4} & -100 & 50 & 10.3182 & 13.0392 & 145.489 & 1098.53 & 4.3896 \\
\hline
\end{tabular}
\end{center}
\caption{Numerical samplings corresponding to the Minkowskian solution, i.e. $\langle V_0 \rangle = 0$. Other flux parameters are set as: $e_0 = 0, \, m^1 =0, \, m_0 = 1, \, {\rm H}^1 =1$, which leads to axions being stabilized as:  $\langle b^1 \rangle= 0$ and $\langle \xi^0 \rangle \, {\rm H}_0 + \langle \xi_1 \rangle\, {\rm H}^{1} = 0$.}
\label{tab_samplings3a}
\end{table}

\noindent
\begin{table}[H]
\begin{center}
\begin{tabular}{|c||c|} 
\hline
Sample &  Eigenvalues of $\langle V_{ij} \rangle.10^6$ \\
\hline
\hline
\bf{S1} &  \{0.531391, 0.23274, 0.104622, 0.000267061, 0.0000988211, 0\} \\
\bf{S2} &  \{0.0697052 , 0.0307571 , 0.0112366 , 0.0000127765 , 4.724$.10^{-6}$ , 0 \} \\
 \bf{S3} & \{0.0175195 , 0.00689047 , 0.00230835 , 1.478$.10^{-6}$ , 5.462$.10^{-7}$ , 0 \} \\
 \bf{S4} &  \{0.09601 , 0.0242602 , 0.0125822 , 0.00135235 , 0.0000172598 , 0 \} \\
\hline
\end{tabular}
\end{center}
\caption{Hessian Eigenvalues for the Minkowskian solution, i.e. $\langle V_0 \rangle = 0$ corresponding to the flux samplings of Table \ref{tab_samplings3a}. This shows that there are no tachyons present, though there is an axionic combination still remaining flat.}
\label{tab_samplings4a}
\end{table}

\subsubsection*{Numerical samplings for de-Sitter solutions:}
We have learnt from the numerical analysis done so far that for slightly larger values of the uplifting parameter $\alpha_4$ as compared to the Minkowskian value mentioned in Eq.~(\ref{eq:alphai's-num}), one can realize tachyon-free dS solutions. For illustration purpose we take the flux sampling {\bf S4} and show the details on dS uplifting starting from AdS solutions via crossing the Minkowskian value, by simply varying the $\alpha_4$ parameter. This is presented in Table \ref{tab_dS-simpling}.
\noindent
\begin{table}[H]
\begin{center}
\begin{tabular}{|c||c|c||c|c|c||c|c||} 
\hline
Model & $\alpha_4$  & $\langle V \rangle.10^9$ & $\langle\rho\rangle$ & $\langle\sigma\rangle$ & $\langle e^{-D}\rangle$ & $\langle{\cal V}\rangle$ & $\langle e^{-\varphi}\rangle$ \\
\hline
\bf{S4} & 0.42 & -3.68518 & 10.2449 & 13.1564 & 141.826 &  1075.28 & 4.32508  \\
\bf{S4} & 0.4242 & 0.026406 & 10.3188 & 13.0384 & 145.518 & 1098.71 & 4.39011  \\
\bf{S4} & 0.425 & 0.700693 & 10.3339 & 13.0152 & 146.278 & 1103.55 & 4.40335  \\
\bf{S4} & 0.43 & 4.67438 & 10.4382 & 12.8638 & 151.559 & 1137.29 & 4.49412 \\
\hline
\end{tabular}
\end{center}
\caption{The numerical samplings for AdS and dS (via crossing the Minkowskian) solutions for a range of values for the $\alpha_4$ parameter. The flux parameters for model {\bf S4} are: $e_1 = -100, \, e_0 = 0, \, m^1 =0, \, m_0 = 1, \,{\rm H}_0 = 50, \, {\rm H}^1 =1$. The plots showing the uplift are presented in Figure \ref{fig_fig1}.}
\label{tab_dS-simpling}
\end{table}

\begin{figure}[h!]
	\begin{center}
		\hspace*{-0.3cm} \includegraphics[width=15.0cm,height=10.5cm]{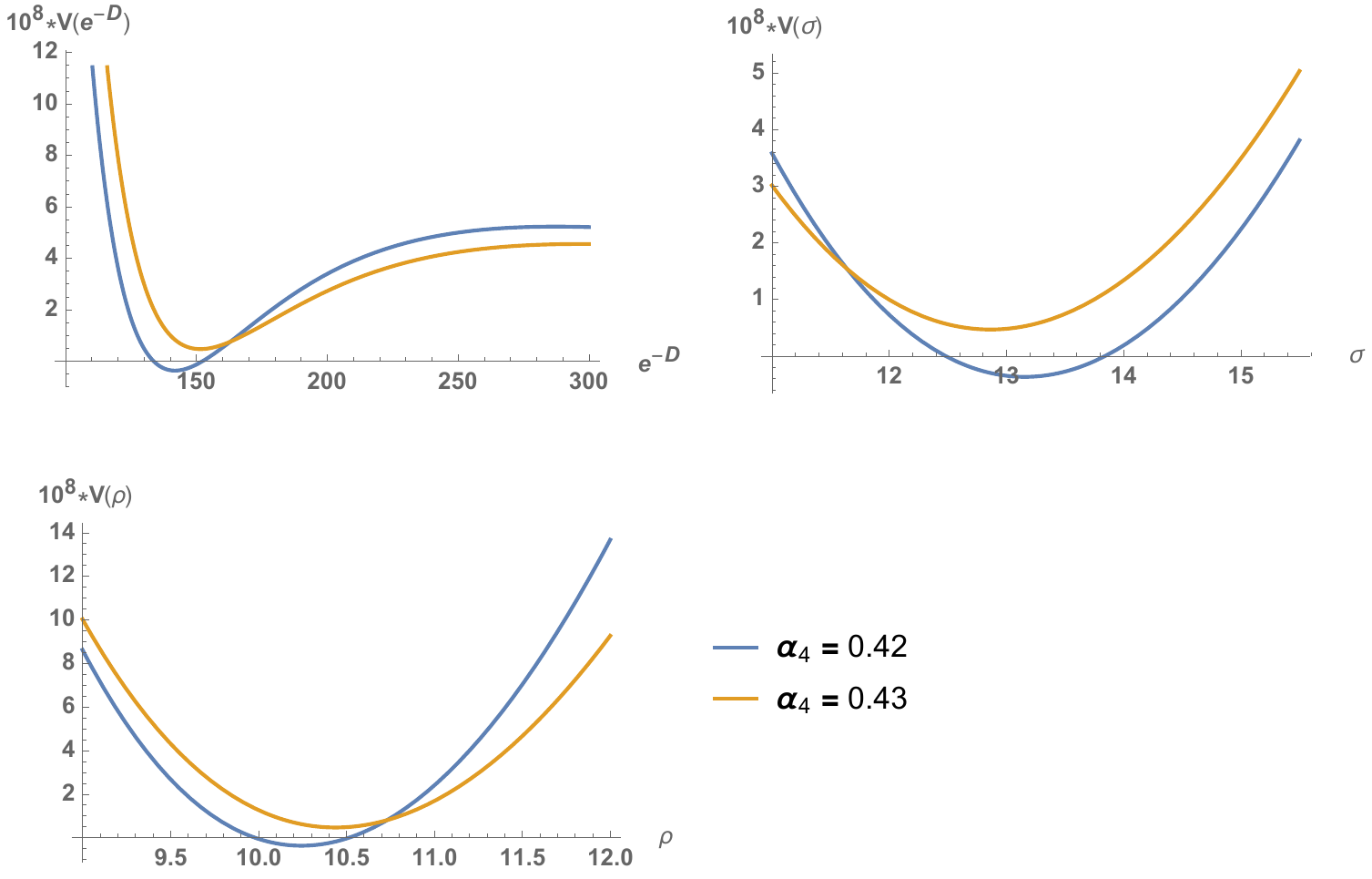}
	\end{center}
\caption{One dimensional slices of scalar potential showing AdS to dS uplifting by $\alpha_4$ parameter.}
\label{fig_fig1}
\end{figure}

\subsection{Comments on viability of the dS vacua}
In this section we present some comments about the viability of the tachyon-free dS solutions we have obtained, and try to explore under which conditions their stability/existence could be questioned, though we do not anticipate such possibilities to occur generically, in the sense of creating issues for the class of AdS/dS solutions we have presented.

\subsubsection*{On the integrality of the fluxes}
Let us recall that our AdS solutions are realized with integer valued fluxes while in order to get the Minkowskian/dS solution, we need to find fluxes such that one satisfies the following condition according to the $\alpha_4$ parameter defined in Eq. (\ref{eq:alpha4}),
\bea
\label{eq:alpha4-bound}
& & \alpha_4 = \frac{(-e_1)\, \hat{w}_{11}^2}{\hat{\kappa}_{111}\,m_0 \,{\rm H}_0\, {\rm H}^1} \geq 0.424169...
\eea
Note that the equality corresponds to the Minkowskian solution for which $\alpha_4$ needs to satisfy Eq. (\ref{eq:alpha4-Minkowskian}) and this generically leads to an irrational value of $\alpha_4$. Therefore, for integral values of fluxes and the triple intersection number $\hat\kappa_{111}$, a truly Minkowskian solution cannot be achieved. However for de-Sitter solutions one would need to take slightly larger values for $\alpha_4$ as compared to the bound given in Eq.~(\ref{eq:alpha4-bound}). For a chosen value of the $\alpha_4$ parameter, we tabulate the positive real values for $\{\alpha_1, \alpha_2, \alpha_3\}$ obtained as the possible solutions to the extremization conditions which are presented in Table \ref{tab_alphai-for-alpha4}. 

\noindent
\begin{table}[h!]
\begin{center}
\begin{tabular}{|c||c|c|c|c|c|}
\hline
 $\alpha_4$ & $\alpha_1$ & $\alpha_2$ & $\alpha_3$ & $\gamma_0$ & Vacua type \\
\hline
\hline
 0.42 & 0.377135 & 1.02449 & 1.8606 & -0.0737035 & Stable AdS \\
 0.42 & 0.743727 & 1.2576 & 1.60788 & 0.171904 & Tachyonic dS \\
 & & & & & \\
 0.4242 & 0.386954 & 1.03188 & 1.8439 & 0.000528121 & Stable dS \\
 0.4242 & 0.717112 & 1.24318 & 1.61508 & 0.184542 & Tachyonic dS \\
 & & & & & \\
 0.425 & 0.388974 & 1.03339 & 1.84063 & 0.0140139 & Stable dS \\
 0.425 & 0.711939 & 1.24034 & 1.61656 & 0.187088 & Tachyonic dS \\
 & & & & & \\
 0.43 & 0.403017 & 1.04382 & 1.81921 & 0.0934877 & Stable dS \\
 0.43 & 0.678523 & 1.2217 & 1.62688 & 0.204158 & Tachyonic dS \\
\hline
\end{tabular}
\end{center}
\caption{For a chosen value of the (uplifting) parameter $\alpha_4$, the set of positive real solutions for the other three parameters $\{\alpha_1, \alpha_2, \alpha_3\}$ are obtained by solving the three extremization conditions in Eq.~(\ref{eq:Vext-STU}). The parameter determining the sign of potential at a given extremum is defined in Eq.~(\ref{eq:gamma0}). The Minkowskian solution defined by Eq.~(\ref{eq:alphai's-num}) lies in the middle of the table.}
\label{tab_alphai-for-alpha4}
\end{table}

\noindent
In Table \ref{tab_alphai-for-alpha4} we have also introduced a new parameter $\gamma_0$ which determines the sign of the VEV of the scalar potential at a given extremum following from Eq.~(\ref{eq:potVEV}). This parameter $\gamma_0$ is subsequently defined as below:

\bea
& & \hskip-1.5cm \langle V \rangle = \gamma_0\, \frac{(m_0)^\frac52\, ({\rm H}_0)\, ({\rm H}^1)^3}{(-e_1)^\frac92},
\eea
where
\bea
\label{eq:gamma0}
& & \gamma_0 = \frac{1}{12\,\alpha_1^4 \alpha_2^3 \alpha_3^3} \biggl[6 \alpha _1 \alpha _3^{3/2} \left(1-3 \alpha _3^2\right) \alpha _2^3+\left(3 \alpha _2^4+1\right) \alpha _3^3 \alpha _2^2 \\
& & \hskip1cm +3 \alpha _1^2 \left(\alpha_2^2 \alpha _4 \alpha _3^6+\left(2 \alpha _4 \alpha _2^2+3\right) \alpha _3^4+\alpha _2^2 \alpha _4 \alpha _3^2+1\right) \biggr]\,. \nonumber
\eea
Using the set of values for $\alpha_i$ parameters from Table \ref{tab_alphai-for-alpha4} one can easily determine the VEVs of saxions for a chosen flux values using Eq.~(\ref{eq:ansatz-dS}). Moreover, a detailed numerical analysis shows that for $\alpha_4 \geq 0.45$, the three extremization conditions in Eq.~(\ref{eq:Vext-STU}) do not result in positive real solutions for the set of parameters $\{\alpha_1, \alpha_2, \alpha_3\}$ which are used for determining the VEVs of their respective saxions. Given that one needs to satisfy the bound in Eq.~(\ref{eq:alpha4-bound}) this shows that there is only a narrow width for $\alpha_4$ parameter which is available for having the dS solutions. This can be estimated as:
\bea
\label{eq:alpha4-dS-bound}
& & 0.42417 \leq \alpha_4 = \frac{(-e_1)\, \hat{w}_{11}^2}{\hat{\kappa}_{111}\,m_0 \,{\rm H}_0\, {\rm H}^1} \leq 0.44.
\eea
Note that each of the quantities defining the parameter $\alpha_4$ as seen above, are either fluxes or triple intersection numbers and hence should usually take integral values. Moreover as we have already argued to set $m_0 = 1$ and ${\rm H}^1 =1$ for having large VEVs for volume modulus and complex-structure modulus respectively, we have the following condition on the remaining quantities:
\bea
\label{eq:alpha4-dS-bound-1}
& & 0.42417 \leq \alpha_4 = \frac{(-e_1)\, \hat{w}_{11}^2}{\hat{\kappa}_{111} \,{\rm H}_0}\, \leq 0.44.
\eea
Note that increasing $|e_1|$ to have larger volume will demand either choosing quite (unnaturally) large value of intersection number $\hat{\kappa}_{111}$ or a large value of ${\rm H}_0$ flux in order to stay withing the required range of the uplifting parameter $\alpha_4$. However, let us not forget that ${\rm H}_0$ flux enters in the tadpole relation and hence can be bounded, unlike the $|e_1|$ flux. So one may not have much freedom to enlarge ${\rm H}_0$ flux to a high value, though one cannot deny to have this possibility in generic CY orientifold models.

Now this suggests that there is a need for a delicate choice of fluxes as $|e_1|$ large is needed for large volume VEV and a couple of such flux samplings with $\alpha_4 = 0.44$, covering a bit the extreme possibilities can be taken as below,
\bea
\label{eq:two-models}
& {\bf S5:} & \quad e_1 = -44, \quad \hat{w}_{11} = 1, \quad \hat{\kappa}_{111} = 20, \quad {\rm H}_0 = 5, \\
& {\bf S6:} & \quad e_1 = -44, \quad \hat{w}_{11} = 1, \quad \hat{\kappa}_{111} = 10, \quad {\rm H}_0 = 10. \nonumber
\eea

\noindent
\begin{table}[h!]
\begin{center}
\begin{tabular}{|c||c|c||c|c|c||c|c|c||} 
\hline
IIA & $\alpha_4$  & $\langle V \rangle.10^8$ & $\langle\rho\rangle$ & $\langle\sigma\rangle$ & $\langle e^{-D}\rangle$ & $\langle{\cal V}\rangle$ & $\langle e^{-\varphi}\rangle$ & Vacua type\\
\hline
\hline
\bf{S5} & 0.44 & 4.54497 & 7.12593 & 3.9504 & 86.8912 & 361.847 & 4.56786 & Stable dS \\
\bf{S5} & 0.44 & 4.95734 & 7.79555 & 3.70697 & 116.96 & 473.741 & 5.37363  & Tachyonic dS\\
\hline
\bf{S6} & 0.44 & 9.08994 & 7.12593 & 5.58671 & 73.0665 & 361.847 & 3.8411 & Stable dS \\
\bf{S6} & 0.44 & 9.91469 & 7.79555 & 5.24245 & 98.3513 & 473.741 & 4.51866 & Tachyonic dS \\
\hline
\end{tabular}
\end{center}
\caption{Numerical samplings for the two new benchmark model defined in Eq.~(\ref{eq:two-models}) which give stable and tachyonic dS solutions for integral valued fluxes and the triple intersection numbers.}
\label{tab_dS-simpling-111}
\end{table}

\subsubsection*{Scale separation arguments}
Let us note that our flux choice is such that one of the RR fluxes, namely $|e_1|$, which governs the VEV of the overall volume modulus, does not appear in the tadpole condition at all, and hence it is not restricted by any upper bound. Subsequently, it is possible to realize quite large values for the VEV of the overall volume of the compactifying sixfold, i.e. $\langle {\cal V} \rangle \gg 1$ along with weak string coupling $\langle g_s \rangle \ll 1$ as can be seen from the benchmark models presented in Table \ref{tab_dS-simpling} and Table \ref{tab_dS-simpling-111}. Therefore, it is expected that the string scale $(M_s)$ and the Kaluza-Klein states $(M_{\rm KK})$ will be separated from the AdS/dS scale for appropriate choice of fluxes. 

\subsubsection*{Obstructions from the Bianchi identities}

For type IIA orientifold case, it has been found \cite{Ihl:2007ah,Gao:2018ayp} that the two known formulations of Bianchi identities do not result in equivalent sets of constraints on the fluxes, and there are always some ``missing identities" which can have impact on the vacua realized within the so-called symplectic or cohomology formulation of (non-)geometric fluxes. So it might be possible that in such a simple setting which allows only a few number of moduli, it could get hard to consistently turn-on all the needed (geometric) fluxes.

For our simple type IIA setting we have six non-zero fluxes in the low energy dynamics, namely $\{e_1, m_0, {\rm H}_0, {\rm H}^1, \hat{w}_{11}, \hat{w}_1{}^0\}$ and these are to be used for stabilizing six scalars fields. To be specific, there are three saxions $\{D, \sigma, \rho\}$ and three axions $\{\xi_0, \xi_1, {\rm b}^1\}$. However, as we switch-off the F-term geometric fluxes in order to satisfy the Bianchi identities, and to avoid a possibly negative contribution to the scalar potential, we need to allow their presence only through the positive semidefinite D-term effects. Subsequently we end up in having the cohomology form of the Bianchi identities resulting in just a single non-trivial constraint. This also reduces the number of independent fluxes to five, which can be attributed to the root cause of having one axionic combination still flat in the end. So the observation that there is at least one axion combination unfixed due to BIs, one cannot afford to make any other fluxes to zero, in case some additional constraints on the remaining fluxes arise in an explicit model through the so-called missing Bianchi identities, which is unknown for the moment. Similar observations have been made for the type IIB models as well \cite{Shukla:2016xdy}, which in the case of rigid compactification has resulted in a (partial) restoration of the no-scale structure even in the presence of non-geometric fluxes \cite{Shukla:2019akv}. To investigate the issue of missing Bianchi identities is beyond the scope of the current plan as the main tasks and goals in this work have been limited to follow a balanced approach in the search of finding stable de-Sitter solutions for integer fluxes, and if such a candidate model is found, then to enumerate the possible loopholes for future refinements!


\section{Summary and conclusions}
\label{sec_conclusions}

In this work we have presented some simple and explicit type IIA models for exploring the possibilities of realizing tachyon-free (stable) AdS/dS solutions, and using the dictionary of \cite{Shukla:2019wfo,Shukla:2019dqd}, the $T$-dual counterpart in type IIB setting can be consistently realized. In our type IIA orientifold model, we include the so-called geometric flux along with the usual NS-NS three-form flux ${\rm H}_3$ and the RR $p$-form fluxes ${\rm F}_p$ for $p \in \{0, 2, 4, 6\}$. First we have presented the simple form of the four-dimensional type IIA scalar potential as given in Eqns.~(\ref{eq:typeIIA-genpot3}) which has been subsequently used for exploring the possible scenario that could evade the well known dS no-go theorems of a geometric type IIA setup. We have engineered the flux choice such that:
\begin{itemize}

\item{Given that the de Sitter no-go results of type IIA models following from the volume/dilaton analysis can be evaded by simultaneously including the Romans mass term and geometric fluxes \cite{Hertzberg:2007wc,Flauger:2008ad}, we consider Romans mass term $m_0$ and (some of) the geometric flux to be always non-zero. To be specific, we make only those geometric fluxes non-zero which appears in the D-term effects and exploit the F-term geometric fluxes to satisfy the Bianchi identities so that not to loose a positive semidefinite contribute to the scalar potential.}

\item{As said above, all the known NS-NS Bianchi identities (of the cohomology formulation) are satisfied without nullifying any of the $D$-term fluxes which could be useful for uplifting purpose, given their positive semi-definite nature.}

\item{Some of the RR fluxes which couple to ${\rm H}_3$ flux and the geometric flux in the tadpole relation are set to zero. To be specific, these are the ones following from the two-form potential ($F_2$) and six-form potential ($F_6$) denoted as: $m^a = 0$ and $e_0 = 0$.}

\item{There is one flux $|e_1|$ which does not receive an upper bound from the tadpole relation as it can couple only to non-geometric ${\rm Q}$-flux (which we do not include in the current analysis), and hence this flux $|e_1|$ can facilitate the realization of large volume, large complex-structure and weak string coupling VEVs for the AdS as well as dS solutions we have.}

\item{There is one combination of axions which remains flat, and it may be attributed to making too restrictive choice of fluxes for various aforesaid reasons.}

\end{itemize}

\noindent
To summarise, we have presented a class of type IIA model with stable AdS/dS solutions and have argued about their viability under the so-called missing Bianchi identities \cite{Ihl:2007ah,Gao:2018ayp}, which still remains an open issue to be explored, and can lead to extra constraints which may or may not rule out the de Sitter solutions we have found. However, settling the issue of missing Bianchi identities is beyond the scope of the current plan and we leave that for a future work. 


\section*{Acknowledgments}
I would like to thank Fernando Marchesano, David Prieto and Joan Quirant for useful discussions, and collaboration at the initial stage of this project. 
In addition, I am thankful to Paolo Creminelli, Atish Dabholkar and Fernando Quevedo for their support.


\appendix
\setcounter{equation}{0}


\section{Type IIA scalar potential with geometric flux}
\label{sec_appendix1}

In the absence of any non-geometric fluxes, the geometric type IIA scalar potential can be encoded in a set of terms given as below \cite{Shukla:2019wfo,Shukla:2019akv},
\bea
& & V  \equiv V_{\rm R} + V_{\rm NS} + V_{\rm loc} = \left(V_{f_6} + V_{f_4} + V_{f_2} + V_{f_0}\right) + \left(V_{h} + V_{\omega}\right) + V_{\rm loc}, 
\eea
where
\bea
\label{eq:pot2}
& & V_{f_6} = \frac{e^{4D}}{4\,{\cal V}}\,{\rm f}_0^2,  \qquad V_{f_4} =\, \frac{e^{4D}}{4}\, {\rm f}_a \, \tilde{\cal G}^{ab} \,{\rm f}_b , \qquad V_{f_2} = \frac{e^{4D}}{4\,} \, {\rm f}^a \, \tilde{\cal G}_{ab} \, {\rm f}^b, \qquad  V_{f_0} =\, \frac{e^{4D}}{4}\,{\cal V}\, ({\rm f}^0)^2, \nonumber\\
& & V_{h} = \frac{e^{2D}}{4\,{\cal V}}\biggl[\frac{{\rm h}_0^2}{\cal U} +\, \tilde{\cal G}^{ij}\,{\rm h}_{i0} \, {\rm h}_{j0} + \,\tilde{\cal G}_{\lambda \rho} {\rm h}^\lambda{}_0 \, {\rm h}^\rho{}_0 \biggr], \\
& & V_{\omega} = \frac{e^{2D}}{4\,{\cal V}\,}\biggl[{\rm t}^{a}\, {\rm t}^{b} \left(\frac{{\rm h}_a \, {\rm h}_b}{\cal U} + \, \tilde{\cal G}^{ij}\, {\rm h}_{ai} \, {\rm h}_{bj} \, + \,\tilde{\cal G}_{\lambda \rho}\, {\rm h}_a{}^\lambda\, {\rm h}_b{}^\rho \right) \nonumber\\
& & \hskip1cm + \,\frac{1}{\cal U} \bigl({\rm h}_a -  \frac{k_\lambda}{2}\,{\rm h}_a{}^\lambda \bigr) \, \bigl({\cal V}\,\tilde{{\cal G}}^{ab} -{\rm t}^a {\rm t}^b\bigr) \bigl({\rm h}_b -  \frac{k_\rho}{2}\,{\rm h}_b{}^\rho \bigr) \, \nonumber\\
& & \hskip1cm + \, \frac{1}{\, {\cal U}}\, \left({\cal U} \, \hat{\rm h}_\alpha{}^{0} + {\rmz}^\lambda \, \hat{\rm h}_{\alpha \lambda} \right) {\cal V}\,(\hat\kappa_{a\alpha\beta}\, {\rm t}^a)^{-1} \,\left({\cal U} \, \hat{\rm h}_\beta{}^{0} + {\rmz}^\rho \, \hat{\rm h}_{\beta \rho} \right)\biggr]\,, \nonumber\\
& &  V_{\rm loc} = \frac{e^{3D}}{2\, \sqrt{\cal U}} \left[\left({\rm f}^0 \, {\rm h}_0 - {\rm f}^a\, {\rm h}_a \right) - \left({\rm f}^0\, {\rm h}^\lambda{}_0 - {\rm f}^a\, {\rm h}^\lambda{}_a \right)\, \frac{k_\lambda}{2} \right]. \nonumber
\eea
where the various non-zero ``axionic flux orbits" can be written in the following form,
\bea
\label{eq:axionic-flux-nogo2}
& & \hskip-0.3cm {\rm f}_0  = e_0 + \, {\rm b}^a\, e_a + \frac{1}{2} \, \kappa_{abc} \, {\rm b}^a\, {\rm b}^b \,m^c + \frac{1}{6}\, \kappa_{abc}\,  {\rm b}^a \, {\rm b}^b\, {\rm b}^c \, m_0 \\
& & \hskip0.3cm - \, \xi^{0} \, ({\rm H}_0 + {\rm b}^a \, {w}_{a0}) - \, \xi^k \, ({\rm H}_k + {\rm b}^a \, {w}_{ak}) - {\xi}_\lambda \, ({\rm H}^\lambda + {\rm b}^a \, {w}_{a}{}^\lambda) \,, \nonumber\\
& & \hskip-0.3cm {\rm f}_a = e_a + \, \kappa_{abc} \,  {\rm b}^b \,m^c + \frac{1}{2}\, \kappa_{abc}\,  {\rm b}^b\, {\rm b}^c \, m_0 - \, \xi^{0} \, {w}_{a0} - \, \xi^k \, {w}_{ak} - {\xi}_\lambda \, {w}_a{}^\lambda\,, \nonumber\\
& & \hskip-0.3cm {\rm f}^a = m^a + m_0\,  {\rm b}^a \,, \nonumber\\
& & \hskip-0.3cm {\rm f}^0 = m_0\,, \nonumber\\
& & \nonumber\\
& & \hskip-0.3cm {\rm h}_0 = ({\rm H}_0 + {\rm b}^a \, {w}_{a0}) + {\rmz}^k \,({\rm H}_k + {\rm b}^a \, {w}_{ak}) + \, \frac{1}{2} \, \hat{k}_{\lambda mn} {\rmz}^m {\rmz}^n \, ({\rm H}^\lambda + {\rm b}^a \, {w}_{a}{}^\lambda), \nonumber\\
& & \hskip-0.3cm {\rm h}_{k0} = ({\rm H}_k + {\rm b}^a \, {w}_{ak}) +  \, \hat{k}_{\lambda k n}\, {\rmz}^n \, ({\rm H}^\lambda + {\rm b}^a \, {w}_{a}{}^\lambda), \quad {\rm h}^\lambda{}_0 = ({\rm H}^\lambda + {\rm b}^a \, {w}_{a}{}^\lambda)\,, \nonumber\\
& & \hskip-0.3cm {\rm h}_a = w_{a0} + {\rmz}^k \,w_{ak}  + \, \frac{1}{2} \, \hat{k}_{\lambda mn} {\rmz}^m {\rmz}^n \, w_a{}^\lambda, \quad {\rm h}_{ak} = w_{ak} +  \, \hat{k}_{\lambda k n}\, {\rmz}^n \,w_a{}^\lambda, \quad {\rm h}_a{}^\lambda = w_a{}^\lambda \,,\nonumber\\
& & \hskip-0.3cm \hat{\rm h}_{\alpha\lambda} = \hat{w}_{\alpha \lambda} + \hat{k}_{\lambda km} \, {\rmz}^m \, \hat{w}_\alpha{}^{k} - \frac{1}{2} \hat{k}_{\lambda km}  {\rmz}^k {\rmz}^m \hat{w}_\alpha{}^{0}, \nonumber\\
& & \hskip-0.3cm \hat{\rm h}_\alpha{}^{0} = \hat{w}_\alpha{}^{0}. \nonumber
\eea
This scalar potential can be studied for the searching the stable vacua through minimization of the moduli/axions, however one can consider the two-field volume/dilaton analysis to rule out certain scenarios \cite{Hertzberg:2007wc,Flauger:2008ad}. In a more general analysis, one can also include the complex structure moduli in order to further check the de-Sitter solutions allowed by the volume/dilaton analysis. Let us also note that the fluxes allowed by the orientifold projection will be further constrained by the following Bianchi identities,
\bea
& & {\rm H}^{\lambda} \, \hat{w}_{\alpha\lambda} = {\rm H}_{\hat{k}} \, \hat{w}_\alpha{}^{\hat{k}}, \qquad w_a{}^\lambda \, \hat{w}_{\alpha \lambda} = w_{a \hat{k}} \, \hat{w}_\alpha{}^{\hat{k}}
\eea
We consider the type IIA setup with geometric flux such that fluxes with $k$-indices are absent. This is equivalent to not having any odd-moduli $G^a$ in the dual type IIB theory \cite{Shukla:2019wfo,Shukla:2019dqd}. 

For that purpose, we further introduce two new moduli, namely $\rho$ and $\sigma$ via a redefinition in the overall volume (${\cal V}$) of the Calabi Yau threefold and its mirror volume ${\cal U}$ by considering the two-cycle volume moduli $\rmt^a$ and $\rmz^i$ as 
\bea
& & {\rmt}^a = \rho \, \gamma^a, \quad \implies \quad {\cal V} = \rho^3, \qquad \kappa_{abc}\gamma^a\gamma^b\gamma^c = 6,\\
& & {\rmz}^\lambda = \sigma \, \theta^\lambda, \quad \implies \quad {\cal U} = \sigma^3, \qquad k_{\rho\gamma\delta} \theta^\rho\theta^\gamma\theta^\delta = 6,\nonumber
\eea
where $\gamma^a$'s denote the angular K\"ahler moduli while $\theta^\lambda$'s corresponds to the angular K\"ahler moduli on the mirror Calabi Yau threefold. Now we can extract the volume factor $\rho$ from the K\"ahler moduli space metric and its inverse in the following way, 
\bea
\label{eq:IIAmetric-rho}
& & \hskip-1.5cm \tilde{\cal G}_{ab} = \frac{\kappa_a\, \kappa_b - 4\, {\cal V}\, \kappa_{ab}}{4\,{\cal V}} = \rho \, \tilde{g}_{ab}, \qquad \, \, \tilde{\cal G}^{ab} =  \frac{2\, {\rm t}^a \, {\rm t}^b - 4\, {\cal V}\, \kappa^{ab}}{4\,{\cal V}} = \frac{1}{\rho} \, \tilde{g}^{ab}\,,\\
& & \hskip-1.5cm \tilde{\cal G}_{\lambda\rho} = \frac{k_\lambda\, k_\rho - 4\, {\cal U}\, k_{\lambda\rho}}{4\,{\cal U}} = \sigma \, \tilde{g}_{\lambda\rho}, \qquad \, \tilde{\cal G}^{\lambda\rho} = \frac{2\, {\rm z}^\lambda \, {\rm z}^\rho - 4\, {\cal U}\, k^{\lambda\rho}}{4\, {\cal U}} = \frac{1}{\sigma} \, \tilde{g}^{\lambda\rho}, \nonumber\\
& & \hskip-1.5cm \tilde{\cal G}_{jk} = - \, \hat{k}_{jk} = \sigma \, \tilde{g}_{ij}, \qquad \qquad \qquad \, \, \, \, \tilde {\cal G}^{jk} = -\, \hat{k}^{jk} = \frac1\sigma \, \tilde{g}^{ij} \,. \nonumber
\end{eqnarray}
Here the matrix $\tilde{g}_{ab}$ and its inverse $\tilde{g}^{ab}$ do not depend on $\rho$ modulus. Similarly, the matrices $\tilde{g}_{\lambda\rho}, \, \tilde{g}^{\lambda\rho},\, \tilde{g}_{ij}$ and $\tilde{g}^{ij}$ do not depend on the analogous complex structure modulus $\sigma$. Using these new redefinitions the explicit dependence of the $\rho$ and $\sigma$ moduli can be extracted out from the generic type IIA scalar potential pieces given in eqn. (\ref{eq:pot2}) takes the form given in Eqn.~(\ref{eq:typeIIA-genpot3}), where $A_i$'s are some functions of fluxes and moduli other than volume modulus $\rho$, the complex structure modulus $\sigma$ and the 4-dimensional dilaton $D$. For completeness, the explicit expressions of $A_i$'s are given below,
\bea
\label{eq:typeIIA-genpot4}
& & A_1 = \frac14\, ({\rm f}^0)^2, \quad A_2 = \frac14\, {\rm f}^a \, \tilde{g}_{ab} \, {\rm f}^b, \quad A_3 = \frac14\,  {\rm f}_a \, \tilde{g}^{ab} \,{\rm f}_b, \quad A_4 = \frac14\,{\rm f}_0^2, \\
& & A_5 = \frac14\,{\rm h}_0^2, \quad A_6 = \frac14\, {\rm h}_{i0} \, \tilde{g}^{ij}\, {\rm h}_{j0}, \quad A_7 = \frac14\, {\rm h}^\lambda{}_0 \, \tilde{g}_{\lambda \rho}\, {\rm h}^\rho{}_0, \quad A_8 = \frac14 \, {\rm h}_a \, \tilde{g}^{ab} \, {\rm h}_b, \nonumber\\
& & A_9 = \frac14 \biggl[\left(\gamma^{a}\, \gamma^{b} \tilde{g}^{ij}\, {\rm h}_{ai} \, {\rm h}_{bj}\right) + ({\theta}^\lambda \, \hat{\rm h}_{\alpha \lambda})\, (\hat\kappa_{a\alpha\beta}\, \gamma^a)^{-1} \, (\theta^\rho \, \hat{\rm h}_{\beta \rho}) \nonumber\\
& & \hskip1cm - \frac12 \bigl(\,\tilde{g}^{ab} -\gamma^a \gamma^b\bigr) \left(k_{\lambda\rho\gamma} \theta^\rho\theta^\gamma \, {\rm h}_a{}^\lambda \, {\rm h}_b + k_{\lambda\rho\gamma} \theta^\lambda\theta^\gamma \, {\rm h}_a\, {\rm h}_b{}^\rho\right) \biggr], \nonumber\\
& & A_{10}= \frac14 \biggl[\left(\gamma^{a}\, \gamma^{b}\, \tilde{g}_{\lambda \rho}\, {\rm h}_a{}^\lambda\, {\rm h}_b{}^\rho\right) + \left(\hat{\rm h}_\alpha{}^{0} \, (\theta^\rho \, \hat{\rm h}_{\beta \rho}) +  ({\theta}^\lambda \, \hat{\rm h}_{\alpha \lambda})\,\hat{\rm h}_\beta{}^{0} \right) (\hat\kappa_{a\alpha\beta}\, \gamma^a)^{-1} \, \nonumber\\
& & \hskip1cm + \frac14 \bigl(\,\tilde{g}^{ab} -\gamma^a \gamma^b\bigr) \, ({\rm h}_a{}^\lambda)\, ({\rm h}_b{}^\rho)\, (k_{\lambda\rho\gamma} \theta^\rho\theta^\gamma) \, (k_{\lambda\rho\gamma} \theta^\lambda\theta^\gamma) \biggr], \nonumber\\
& & A_{11} = \frac14 \,(\hat{\rm h}_\alpha{}^{0})\ (\hat\kappa_{a\alpha\beta}\, \gamma^a)^{-1} \, (\hat{\rm h}_\beta{}^{0}), \nonumber\\
& & A_{12} = \frac12 \left({\rm f}^0 \, {\rm h}_0 - {\rm f}^a\, {\rm h}_a \right), \quad A_{13} = -\frac14 \, (k_{\lambda\rho\gamma} \theta^\rho\theta^\gamma)\,\left({\rm f}^0\, {\rm h}^\lambda{}_0 - {\rm f}^a\, {\rm h}^\lambda{}_a \right). \nonumber
\eea
This shows that all the $A_i$'s except $A_9,\, A_{10}, \, A_{12}$ and $A_{13}$ are positive semi-definite.




\bibliographystyle{utphys}
\bibliography{reference}

\end{document}